\documentclass[journal=jctcce,manuscript=article]{achemso}

\usepackage{chemformula} 
\usepackage[T1]{fontenc} 
\usepackage{siunitx}
\usepackage[version=3]{mhchem}
\usepackage[flushleft]{threeparttable}
\usepackage{hyperref}
\usepackage{adjustbox}
\usepackage{physics}

\author{Christian S. Ahart}
\affiliation[Imperial College London]
{Imperial College London, Department of Chemistry and Thomas Young Centre, Molecular Sciences Research Hub, London W12 0BZ, UK}
\email{c.ahart@imperial.ac.uk}
\author{Sergey K. Chulkov}
\affiliation[University of Lincoln]
{University of Lincoln, School of Mathematics and Physics, Lincoln LN6 7TS, UK}
\author{Clotilde S. Cucinotta}
\affiliation[Imperial College London]
{Imperial College London, Department of Chemistry and Thomas Young Centre, Molecular Sciences Research Hub, London W12 0BZ, UK}
\email{c.cucinotta@imperial.ac.uk}

\title[An \textsf{achemso} demo]
{Enabling Ab-Initio Molecular Dynamics under Bias: The CP2K+SMEAGOL Interface for Integrating Density Functional Theory and Non-Equilibrium Green Functions}


\begin{document}

\begin{tocentry}
  \includegraphics[width=0.7\columnwidth]{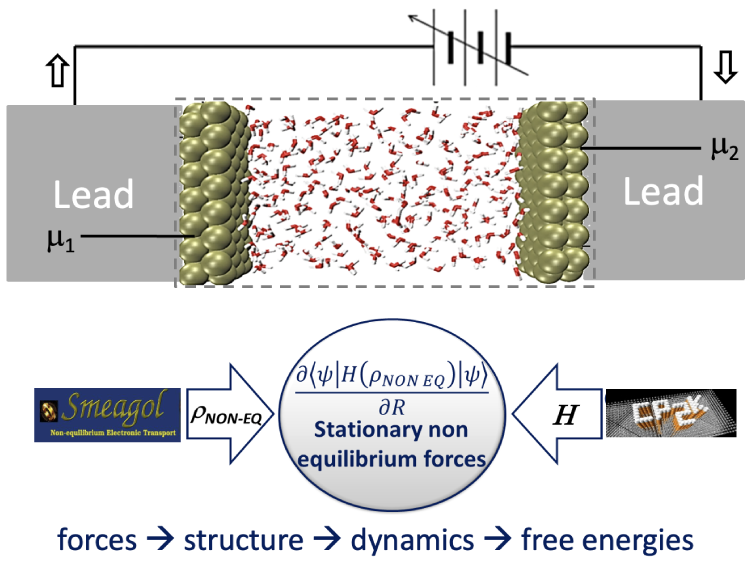}
\end{tocentry}

\clearpage
\begin{abstract}
Density functional theory (DFT) combined with non-equilibrium Green’s functions (NEGF) is a powerful approach to model quantum transport under external bias potentials, at reasonable computational cost. In this work we present a new interface between the popular mixed Gaussian/plane wave electronic structure package CP2K and the NEGF code SMEAGOL, the most feature-rich implementation of DFT-NEGF available for CP2K to-date. The CP2K+SMEAGOL interface includes the implementation of current induced forces. We verify this implementation for a variety of systems: an infinite 1D Au wire, a parallel-plate capacitor and a Au-H2-Au junction. We find good agreement with SMEAGOL calculations performed with SIESTA for the same systems, and with the example of a solvated Au wire demonstrate for the first time that DFT-NEGF can be used to perform molecular dynamics simulations under bias of large-scale condensed phase systems under realistic operating conditions.
\end{abstract}

\section{Introduction} \label{section:introduction}

Improving the performance of electro-catalytic reactions such as in batteries, solar cells and capacitors is essential to address growing energy demands and costs. While much progress has been made, there remains a significant gap between the theoretical understanding of microscopic phenomena and the macroscopic outcomes of experiments\cite{nielsenFirstPrinciplesModeling2015}. Simulation methods such density functional theory (DFT) can provide valuable information about the structure and dynamical properties of electrochemical (EC) systems\cite{sundararamanImprovingAccuracyAtomistic2022,Khatib2021}, however conventional methodologies do not allow for the study of systems under realistic operating conditions such as under applied potential and current flow.
Indeed, these methodologies are implemented within a canonical framework, but EC transformation is intrinsically grand canonical. To describe the natural environment of EC transformation, we need to perform molecular dynamics under bias and have an explicit open-boundary description of the electrons, which must be free to enter and leave the computational cell.

DFT combined with non-equilibrium Green’s functions (NEGF) is a powerful approach to model quantum transport under external bias potentials\cite{xueFirstprinciplesBasedMatrix2002, taylorInitioModelingQuantum2001}. In a standard DFT-NEGF calculation the non-equilibrium density is calculated self-consistently by integrating a Green's function that combines the DFT Hamiltonian with semi-infinite electrodes, including the effect of an applied potential and current. Experimental transport properties such as current-voltage (I-V) curves and conductance measurements have been shown to be accurately reproduced through the DFT-NEGF approach\cite{RevModPhys.92.035001, 10.1039/BK9781849731331-00179,rochaSpinMolecularElectronics2006,pomorskiCapacitanceInducedCharges2004}, at a reasonable computational cost.

In recent years there have been many new implementations of DFT-NEGF in popular DFT packages\cite{rochaSpinMolecularElectronics2006,chulkovCP2KElectronTransport, smidstrupQuantumATKIntegratedPlatform2020, papiorImprovementsNonequilibriumTransport2017,bagretsSpinPolarizedElectronTransport2013, chenInitioNonequilibriumQuantum2012, ozakiEfficientImplementationNonequilibrium2010, sahaFirstprinciplesMethodologyQuantum2009, kaminskiTuningConductanceBenzenebased2016, pedrozaBiasdependentLocalStructure2018, Lu2011}, however standard calculations of transport properties are performed at fixed atomic positions and applications to dynamic condensed phase systems remain rare\cite{choiElectronTransportNanoconfined2023}. In this work we present a new interface between the NEGF code SMEAGOL and CP2K, a popular DFT software package optimised for condensed phase molecular dynamics simulations\cite{Kuhne2020}. We demonstrate that it is possible to perform large-scale molecular dynamics simulations under realistic operating conditions using the DFT-NEGF approach, to the best of our knowledge the first such calculations to be performed.

We chose to interface CP2K with SMEAGOL as it is a well established and feature-rich implementation of DFT-NEGF\cite{rochaSpinMolecularElectronics2006}. The existing version of SMEAGOL however is closely bound to the 2003 release of the DFT package SIESTA\cite{solerSIESTAMethodInitio2002}, and as such is lacking many features present in state-of-the-art DFT packages\cite{garciaSIESTARecentDevelopments2020}. To avoid confusion, we refer to this original version of SMEAGOL as SIESTA+SMEAGOL and our new interface as CP2K+SMEAGOL. The philosophy during the development of CP2K+SMEAGOL was to reuse the original SMEAGOL code as much as possible, creating a new standalone SMEAGOL library linkable with any DFT software package. Our new CP2K+SMEAGOL interface is available in the development version of CP2K, and instructions on how to obtain the SMEAGOL library as well as example input files can be found at the \href{https://wiki.ch.ic.ac.uk/wiki/index.php?title=Potential_control_and_current_induced_forces_using_CP2K%2BSMEAGOL}{Imperial College London Nano Electrochemistry Group’s wiki}. 

We note that while there are several DFT-NEGF implementations currently available for CP2K\cite{chulkovCP2KElectronTransport,calderaraPushingBackLimit2015}, CP2K+SMEAGOL is the most feature-rich and complete to-date. The support for DFT-NEGF natively in CP2K is limited to $\Gamma$-point only transport calculations, without forces required for geometry optimisation or molecular dynamics\cite{chulkovCP2KElectronTransport}. 
 
The remainder of the paper is organised as follows. Section \ref{section:theory} summarises briefly the theory of DFT-NEGF and the implementation in CP2K+SMEAGOL, followed by results at both zero bias (Section \ref{section:au_chain}) and finite bias (Section \ref{section:au_capacitor}-\ref{section:au_h2_au}). We also show that CP2K+SMEAGOL can be used to perform molecular dynamics of large solvated systems (Section \ref{section:au_wire_solvated}), and discuss the performance and as well as methodologies that may be employed to accelerate the dynamics in future work (Section \ref{section:performance}). Concluding remarks are made in Section \ref{section:conclusion}. 

\section{Theory and implementation} \label{section:theory}
DFT-NEGF is a well established method, with many recent implementations in popular DFT packages\cite{chulkovCP2KElectronTransport, smidstrupQuantumATKIntegratedPlatform2020, papiorImprovementsNonequilibriumTransport2017,bagretsSpinPolarizedElectronTransport2013, chenInitioNonequilibriumQuantum2012, ozakiEfficientImplementationNonequilibrium2010, sahaFirstprinciplesMethodologyQuantum2009, kaminskiTuningConductanceBenzenebased2016,pedrozaBiasdependentLocalStructure2018}. As such, we choose to only briefly summarise the theory relevant to this work.

\begin{figure}[t!]
         \includegraphics[width=0.6\columnwidth]{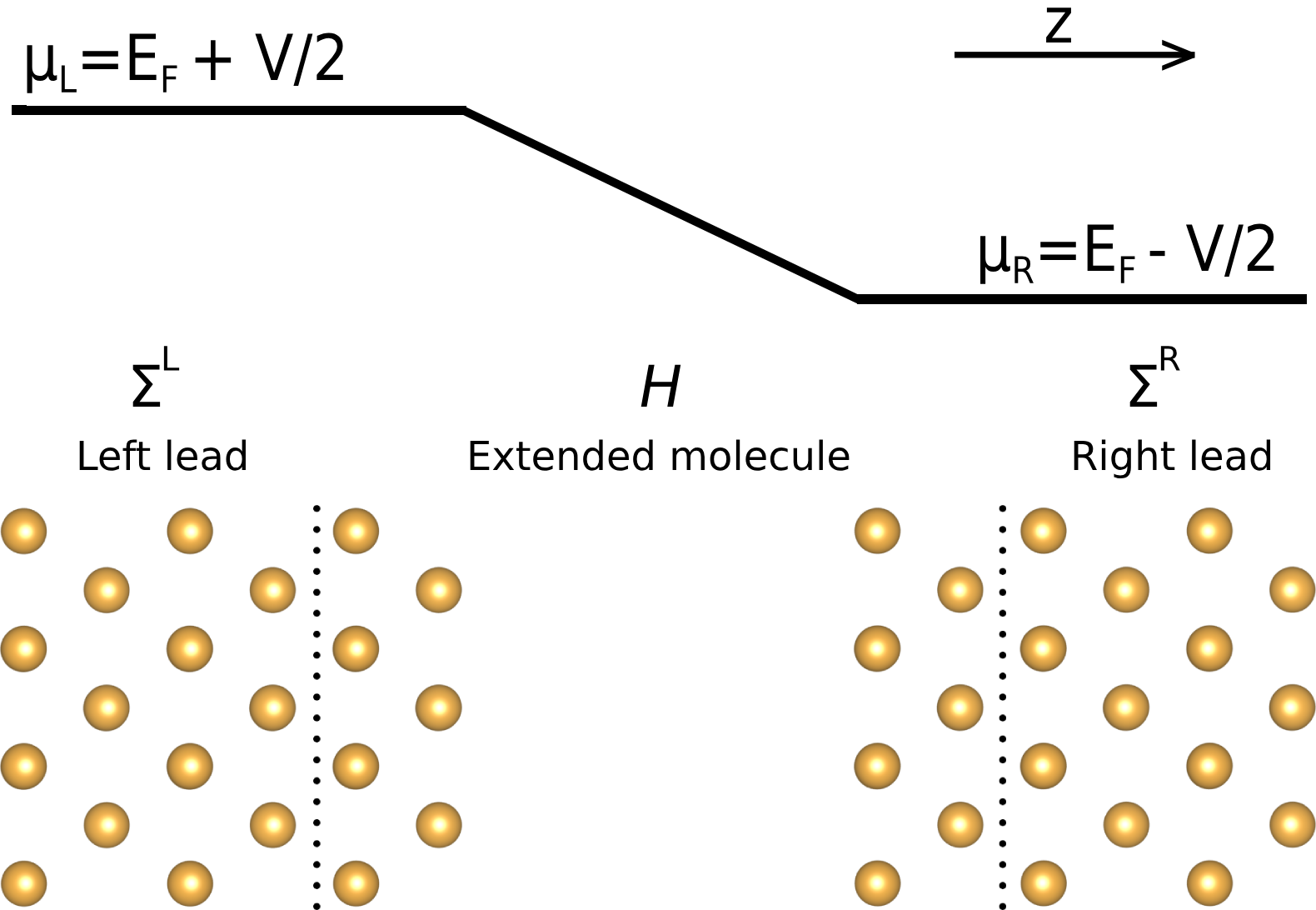}
     \caption{Schematic view of a Au capacitor used for CP2K+SMEAGOL and SIESTA+SMEAGOL calculations.}
     \label{fig:bias_em_capacitor}
\end{figure}

A representation of the typical arrangement used in a DFT-NEGF calculation is shown in Fig. \ref{fig:bias_em_capacitor}, composed of a central region termed the `extended molecule' (EM) attached to two semi-infinite electrodes or `leads'. The two leads are kept at different chemical potential by a battery and are able to exchange electrons with the extended region. 

The retarded Green's function $G(\epsilon)$, referred to as the Green's function in the remainder of this work, is obtained by performing an inversion of the DFT Hamiltonian matrix $H$ combined with the overlap matrix $S$ and the leads self-energies $\Sigma^{\mathrm{L}}$ and $\Sigma^{\mathrm{R}}$\cite{taylorInitioModelingQuantum2001, brandbygeDensityfunctionalMethodNonequilibrium2002},  

\begin{equation}
G(\epsilon)=[(\epsilon+i\delta_{+})S - H-\Sigma^{\mathrm{L}}(\epsilon)-\Sigma^{\mathrm{R}}(\epsilon)]^{-1},
\label{eq:greens_function}
\end{equation}
where $\delta_{+}$ is an infinitesimal positive number. The self energies are in general non-hermitian matrices that contain  information about the electronic structure of the leads and possibly the bias. In principle the leads self-energies $\Sigma^{\mathrm{L}}$ and $\Sigma^{\mathrm{R}}$ can differ, however in practice this is challenging to converge and therefore for all calculations performed in this work the left and right leads are identical. 

Similarly to SIESTA+SMEAGOL, the Hamiltonian and the non-equilibrium density of the central region are calculated self-consistently using the NEGF scheme, with boundary conditions set by self-energies that are static and independent of the charge density in the central region. This assumption holds if the central region is sufficiently large, ensuring that changes in charge density are screened before reaching the boundaries. For the metallic leads typically used, which have a short screening length, this requirement is met within a few atomic layers. In calculations, these layers need to be included at the boundaries of the central region to extend the leads. With the NEGF method, the currents entering and leaving the EM are balanced, allowing to evaluate the electron density in the EM, the electrostatic potential and the current.

The electron density is split into left and right contributions $D = D_{\mathrm{L}} + D_{\mathrm{R}}$, calculated as an integral of the Green's function

\begin{equation}
 D^{\mathrm{L}} = \int \rho^{\mathrm{L}}(\epsilon) f(\frac{\epsilon - \mu_{\mathrm{L}}}{k_{B}T_{\mathrm{L}}}) \mathrm{d} \epsilon, 
\label{eq:total_density_matrix_left}
\end{equation}
where $f$ is the Fermi-Dirac distribution of the left lead described by the chemical potential $\mu_{\mathrm{L}}$, $k_{B}$ the Boltzmann constant, $T_{\mathrm{L}}$ is the electronic temperature of the left lead and $\rho^{\mathrm{L}}(\epsilon)$ is the spectral density matrix 

\begin{equation}
\rho^{\mathrm{L}}(\epsilon) = \frac{1}{2\pi}G(\epsilon)\Gamma^{\mathrm{L}}(\epsilon)G^\dagger(\epsilon),
\label{eq:density_matrix_left}
\end{equation}
calculated from the Green's function and the broadening function $\Gamma^{\mathrm{L}}$ of the left electrode, 

\begin{equation}
\Gamma^{\mathrm{L}}=\frac{1}{\mathbf{i}} \left( \Sigma^{\mathrm{L}}-(\Sigma^{\mathrm{L}})^\dagger \right).
\label{eq:broadening_function_left}
\end{equation}

The calculation of the electron density as an integral of the Green's function across the entire energy space in Eq. \ref{eq:total_density_matrix_left} would be very demanding, and therefore the integral is split further into an equilibrium contribution $D^{\mathrm{L}}_{\mathrm{eq}}$, which can be integrated on a complex contour, and a non-equilibrium contribution $\Delta^{\mathrm{R}}_{\mathrm{neq}}$, which needs to be integrated along the real axis. Notably, the non-equilibrium contribution to the density only needs to be evaluated for energies within the bias window. The total density is therefore written as

\begin{equation}
D = D^{\mathrm{L}}_{\mathrm{eq}}  +\Delta^{\mathrm{R}}_{\mathrm{neq}},
\label{eq:single_contour}
\end{equation}
where 
\begin{equation}
D^{\mathrm{L}}_{\mathrm{eq}}=\frac{1}{\pi} \Im{\int \mathrm{d} \epsilon G(\epsilon) f \left( \frac{\epsilon - \mu_{\mathrm{L}}}{k_{B}T_{\mathrm{L}}}\right) }, 
\label{eq:density_matrix_equilibrium_complex}
\end{equation}
and
\begin{equation}
\Delta^{\mathrm{R}}_{\mathrm{neq}} =\int \mathrm{d} \epsilon\rho^{\mathrm{R}}(\epsilon) \left[ f \left( \frac{\epsilon - \mu_{\mathrm{R}}}{k_{B}T_{\mathrm{R}}}\right) - f \left( \frac{\epsilon - \mu_{\mathrm{L}}}{k_{B}T_{\mathrm{L}}}\right) \right].
\label{eq:density_matrix_nonequilibrium_real}
\end{equation}

We could re-write Eq. \ref{eq:single_contour} as $D =D^{\mathrm{R}}_{\mathrm{eq}}+\Delta^{\mathrm{L}}_{\mathrm{neq}}$, where L and R are exchanged. The original version of SMEAGOL calculates the density matrix using either approach, or as an average of both\cite{rochaSpinMolecularElectronics2006}. This approach however does not correctly occupy bound states within the bias window, discussed further in Section \ref{section:au_capacitor}. As such we have updated SMEAGOL to more closely follow the work of Brandbyge et al.\cite{brandbygeDensityfunctionalMethodNonequilibrium2002}, calculating the density matrix as a weighted sum of the two integrals

\begin{equation}
D_{ij} = W_{ij} \left[ D^{\mathrm{L}}_{ij\mathrm{, eq}}  +\Delta^{\mathrm{R}}_{ij\mathrm{, neq}} \right] + (1-W_{ij}) \left[ D^{\mathrm{R}}_{ij\mathrm{, eq}}  +\Delta^{\mathrm{L}}_{ij\mathrm{, neq}}  \right],
\label{eq:density_matrix_double_contour}
\end{equation}
where
\begin{equation}
W_{ij} = \frac{(\Delta^{\mathrm{L}}_{ij\mathrm{, neq}})^{2}}{(\Delta^{\mathrm{L}}_{ij\mathrm{, neq}})^{2}+(\Delta^{\mathrm{R}}_{ij\mathrm{, neq}})^{2}}.
\label{eq:double_contour_weight}
\end{equation}

Following the self-consistent calculation of the non-equilibrium density a number of transport quantities can be calculated using the DFT-NEGF approach, such as energy dependent total transmission coefficient\cite{buttikerGeneralizedManychannelConductance1985, rochaSpinMolecularElectronics2006}
\begin{equation}
T(\epsilon) = \Tr{ G(\epsilon) \Gamma^{\mathrm{L}} G^{\dagger}(\epsilon) \Gamma^{\mathrm{L}} },
\label{eq:transmission}
\end{equation}
and the current, calculated using the Landauer-Buttiker formalism\cite{meirLandauerFormulaCurrent1992}
\begin{equation}
I = \frac{e}{h} \int \mathrm{d} \epsilon T(\epsilon) \left[ f \left( \frac{\epsilon - \mu_{\mathrm{L}}}{k_{B}T_{\mathrm{L}}}\right) - f \left( \frac{\epsilon - \mu_{\mathrm{R}}}{k_{B}T_{\mathrm{R}}}\right) \right].
\label{eq:current}
\end{equation}

An electric current can significantly change the forces experienced by a nano-system, and therefore its dynamics. This is at the origin of electromigration phenomena and of the rupture of nanowires under bias conditions. These forces are non-conservative\cite{todorovCurrentinducedForcesSimple2014}, thus the way the electric current affects the dynamics of an atom can vary along the trajectory of that atom, causing instabilities. The forces are calculated via the time derivative of the atomic momenta, resulting from the average of the gradients (with respect to atomic positions) of the full Hamiltonian, according to the expression:
\begin{equation}
\mathbf{F}_i = \frac{d}{dt} \langle \Psi(t) | \hat{P}_i | \Psi(t) \rangle =- \langle \Psi(t) | \nabla_{\mathbf{R}_i} \hat{H} | \Psi(t) \rangle 
\label{eq:force}
\end{equation}
The Hamiltonian in this expression is calculated using the non-equilibrium density obtained by solving the NEGF problem, which embed instantaneously the information about the stationary charge redistribution induced by current and bias. We use these non-conservative forces to perform Born-Oppenheimer molecular dynamics under bias and currents. We would like to highlight that this expression for the force acting on ions is generally different from the Hellmann-Feynman expression, $\mathbf{F}_i = -\nabla_{\mathbf{R}_i}\expval{\hat{H}}{\Psi}$. However, according to Rungger et al. \cite{zhangCurrentinducedEnergyBarrier2011} and Di Ventra et al. \cite{diventraHellmannFeynmanTheoremDefinition2000} if the state is stationary (with the only time dependence being through the exponential term) and is square integrable, the two expressions become formally identical. 

Another consideration for any DFT-NEGF implementation is to correctly express in term of the Green functions, the force term arising from the derivative of the density matrix, $D_{ij}$\cite{pulayInitioCalculationForce1969, Vandevondele2005},

\begin{equation}
\sum_{ij} (\nabla_I D_{ij}) K_{ij} = -2\sum_{ij}\Omega_{ij}\bra{\nabla_I \phi_i}\ket{\phi_{j}},
\end{equation}
where, I is the ion index, $K_{ij}$ is the Kohn-Sham matrix, $\Omega_{ij}$ is the energy weighted density matrix, and $\phi_{i/j}$ are localised orbital functions. 
Specifically, the $\Omega_{ij}$ matrix is calculated in the same manner as the density matrix (Eq. \ref{eq:total_density_matrix_left}), with an additional energy term $\epsilon$ in the integral\cite{zhangCurrentinducedEnergyBarrier2011}.
Notably, this term is often neglected in existing implementations of current-induced forces\cite{pedrozaBiasdependentLocalStructure2018a}, however without it the forces are incorrect  (Section \ref{section:au_chain}). This implementation enables the modelling of the `electron wind' component of the current-induced force, which models the transfer of momentum from electrons to ions under fixed boundary conditions, as discussed by Dundas et al. \cite{dundasCurrentdrivenAtomicWaterwheels2009} and Zhang et al. \cite{zhangCurrentinducedEnergyBarrier2011}. We neglect any further electron-ion coupling, including coupling with electrode phonons. We assume that we are in the limit where the fluctuating forces leading to Joule heating\cite{Lü2011} and the phase change of the electronic wave functions due to atomic motion\cite{doi:10.1021/nl904233u} are not significant.

\section{Results} \label{section:results}

We present validation of CP2K+SMEAGOL at both zero bias (Section \ref{section:au_chain}) and finite bias (Section \ref{section:au_capacitor}-\ref{section:au_h2_au}), for a variety of systems. We also show that CP2K+SMEAGOL can be used to perform molecular dynamics of large solvated systems (Section \ref{section:au_wire_solvated}), and discuss the performance and as well as methodologies that may be employed to accelerate the dynamics in future work (Section \ref{section:performance}). 

\subsection{Zero bias forces} \label{section:au_chain}

\begin{figure}[t!]
         \includegraphics[width=0.6\columnwidth]{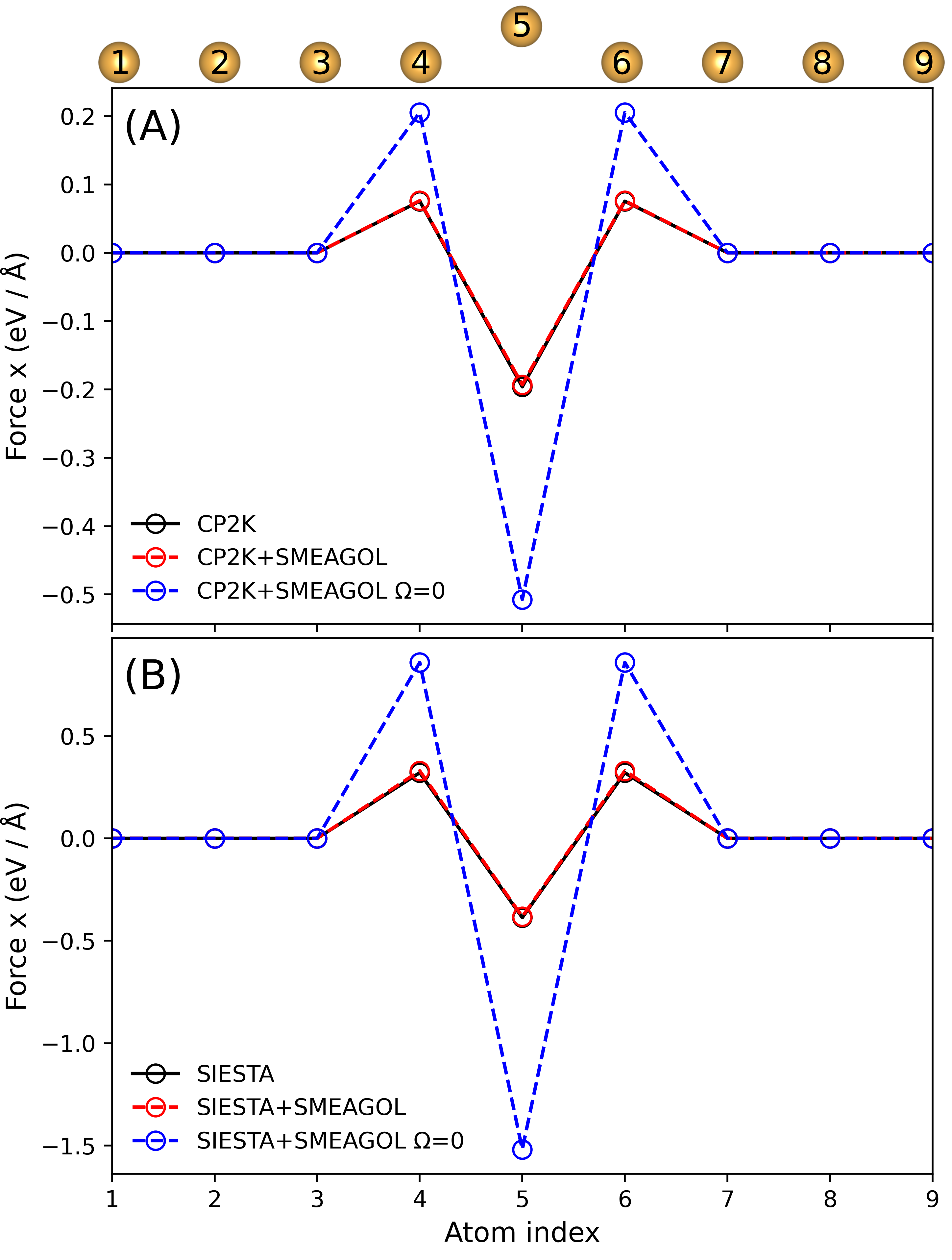}
     \caption{Zero bias tests for an infinite Au wire. (A) Atomic forces calculated with CP2K and CP2K+SMEAGOL. (B) Atomic forces calculated with SIESTA and SIESTA+SMEAGOL. The y and z components of the atomic forces are shown in ESI Fig. 1.}
     \label{fig:au_wire_force_paper}
\end{figure}

A simple validation of our CP2K+SMEAGOL interface is to ensure that the forces calculated at zero bias with CP2K+SMEAGOL are equal to the forces calculated only with CP2K. We use the same system setup as Zhang et al.\cite{zhangCurrentinducedEnergyBarrier2011}, an infinite Au chain composed of 9 atoms. The central Au atom is displaced by 1 Å in the +x direction, such that there is a restoring force acting towards the equilibrium position. The left lead, extended molecule and right lead are each composed of 3 Au atoms. The LDA functional is used, with a single-$\zeta$ basis set for the Au(6s, 5d) electrons. The same system setup is used to perform reference calculations in SIESTA+SMEAGOL, using an explicit cutoff radius for Au(6s) of 6.5 Bohr and Au(5d) of 5.85 Bohr. Unless otherwise stated, the same computational setup is used for calculations in this work.

Fig. \ref{fig:au_wire_force_paper} shows the x component of the force calculated with both CP2K and CP2K+SMEAGOL at zero bias, as well as reference calculations with SIESTA and SIESTA+SMEAGOL. While the exact value of the force differs between CP2K and SIESTA, they are both consistent with the forces calculated with SMEAGOL at zero bias. We also demonstrate the effect of neglecting the forces calculated from the energy density matrix term $\Omega$, as performed in some DFT-NEGF codes\cite{pedrozaBiasdependentLocalStructure2018a}, resulting in qualitatively incorrect forces.

\subsection{Parallel-plate capacitor} \label{section:au_capacitor}

\begin{figure}[t!]
         \includegraphics[width=1\columnwidth]{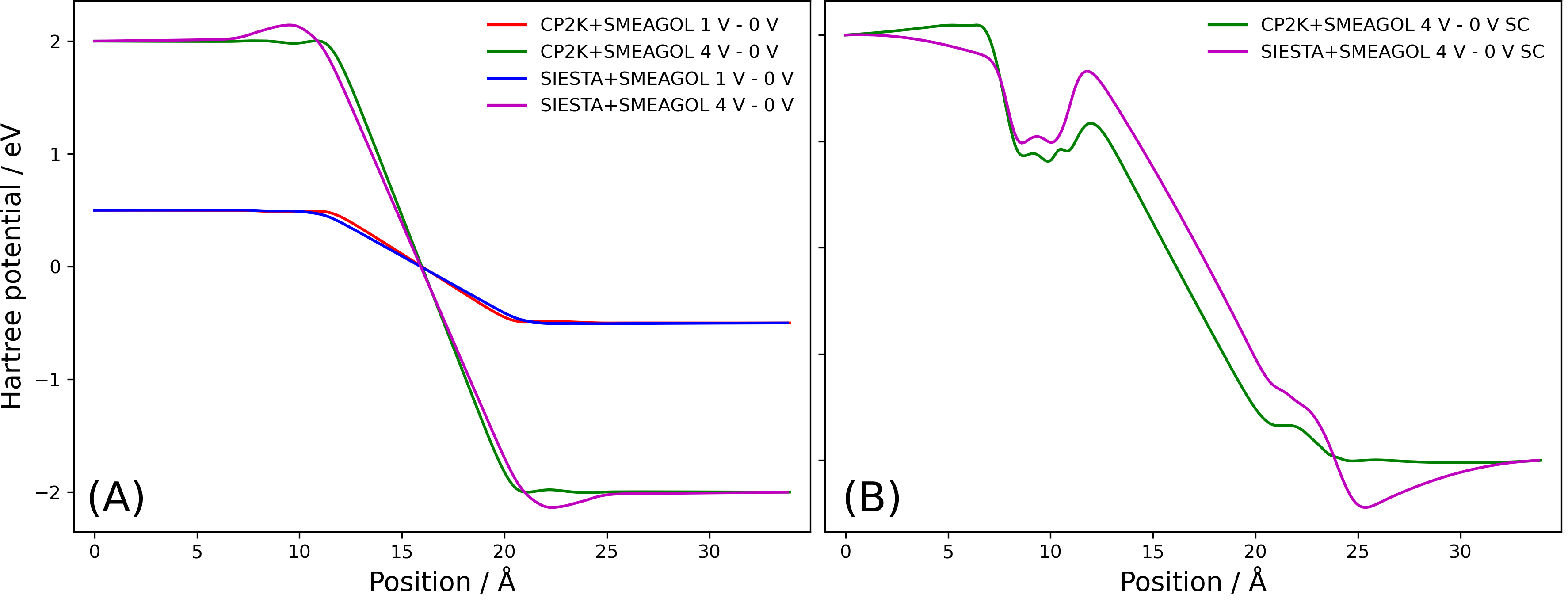}
     \caption{Finite bias tests for a parallel-plate capacitor. (A) Planar average of the Hartree potential difference calculated with and without an applied bias using CP2K+SMEAGOL and SIESTA+SMEAGOL. (B) Planar average of the Hartree potential difference calculated with and without an applied bias using a single contour evaluation of the Green's function, leading to an incorrect occupation of bound states within the bias window. The corresponding planar average of the charge density is shown in ESI Fig. 2.}
     \label{fig:au-capacitor_paper}
\end{figure}

Another key validation of our CP2K+SMEAGOL interface is to reproduce the potential drop across a parallel-plate capacitor, a popular benchmark system for DFT-NEGF implementations\cite{rochaSpinMolecularElectronics2006, brandbygeDensityfunctionalMethodNonequilibrium2002}. The structure used is shown in Fig. \ref{fig:bias_em_capacitor}, composed of two symmetric 2x2 Au(001) slabs each with 6 Au layers, separated by 12 Å vacuum. The left and the right leads are each composed of 4 Au layers, necessary to reproduce the ABAB periodicity of the Au(001) surface in the semi-infinite leads. 

Fig. \ref{fig:au-capacitor_paper} shows the planar average of the Hartree potential difference across the parallel-plate capacitor with and without an applied bias. The CP2K+SMEAGOL and SIESTA+SMEAGOL results are consistent, demonstrating a constant potential in the leads and a linear drop in the vacuum region. The difference in the charge accumulation at the surface of the electrodes can be attributed to the difference in the basis sets used for the CP2K and SIESTA calculations. 

Fig. \ref{fig:au-capacitor_paper} also demonstrates that the Hartree potential calculated using a single contour evaluation of the Green's function (Eq. \ref{eq:single_contour}) is qualitatively incorrect at high bias, with asymmetric charge density at the left and right electrodes. Using our newly implemented weighted double contour (Eq. \ref{eq:density_matrix_double_contour}) scheme in SMEAGOL we produce a qualitatively correct Hartree potential, with symmetric charge density on the left and right electrode.  Brandbyge et al.\cite{brandbygeDensityfunctionalMethodNonequilibrium2002} attributed this difference to the presence of bound states within the bias window that only couple to one electrode, and therefore a single contour evaluation does not correctly populate these states. As the use of the weighted double contour is only necessary at high applied bias, generally above 2 V in our experience, we only use a single contour evaluation of the Green's function for all other calculations in this work.

\subsection{Geometry optimisation of an Au-H$_{2}$-Au junction} \label{section:au_h2_au}

\begin{figure}[t!]
         \includegraphics[width=0.6\columnwidth]{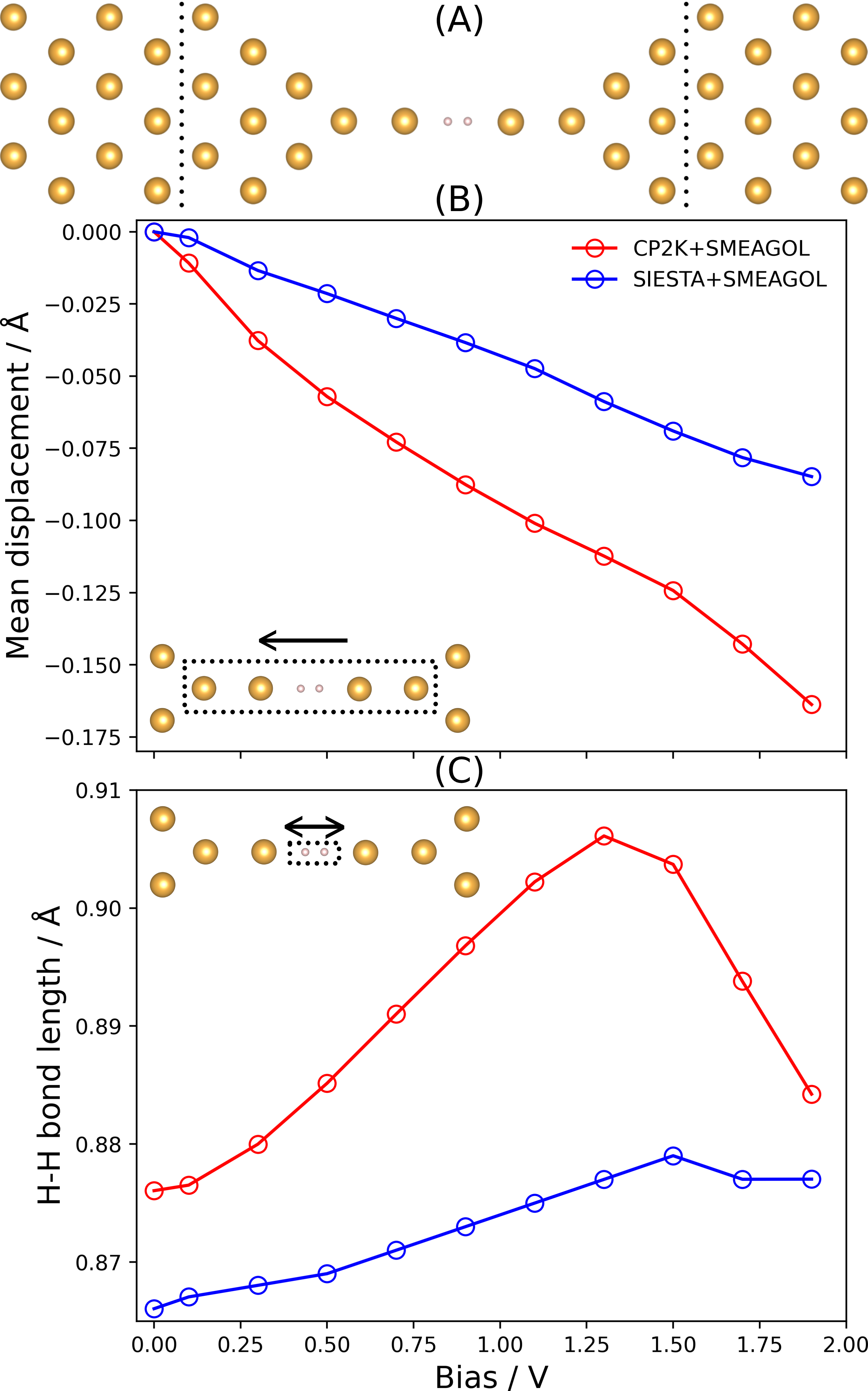}
     \caption{Finite bias geometry optimisation of a Au-H$_{2}$-Au junction. (A) Optimised atomic structure of the Au-H$_{2}$-Au junction, where yellow and white spheres represent Au and H atoms respectively. (B) Mean displacement of the six highlighted atoms as a function of the applied bias. (C) Change in the H-H bond length as a function of the applied bias.}
     \label{fig:au-h2}
\end{figure}

Hydrogen molecules have been shown to strongly interact with Au nanojunctions\cite{csonkaPullingGoldNanowires2006}, and as such the resulting structures and their transport properties have received much interest both experimentally\cite{csonkaFractionalConductanceHydrogenEmbedded2003,csonkaPullingGoldNanowires2006} and computationally\cite{barnettHydrogenWeldingHydrogen2004,jelinekHydrogenDissociationAu2006,baiCurrentinducedPhononRenormalization2016,frederiksenInelasticFingerprintsHydrogen2007,frederiksenInelasticFingerprintsHydrogen2007}. Having demonstrated that CP2K+SMEAGOL can reproduce forces and charge density consistent with SIESTA+SMEAGOL, we perform calculations for the Au-H$_{2}$-Au junction based on the work of Bai et al.\cite{baiCurrentinducedPhononRenormalization2016} performed using SIESTA+SMEAGOL.

The structure used is shown in Fig. \ref{fig:au-h2}, composed of a hydrogen molecule sandwiched between two 3x3 Au(100) slabs with 7 layers (left) and 6 layers (right). All atoms are constrained in the system except for the 2 H atoms and the 6 Au atoms either side. The PBE functional is used, with a single-$\zeta$ basis set for the Au(6s, 5d) electrons and a double-$\zeta$ basis set for the H(2s) electrons plus polarisation functions. For the reference SIESTA+SMEAGOL calculations we use an explicit cutoff radius for Au(6s) of 6 Bohr, Au(5d) of 5.5 Bohr, H(2s) of 5.5 Bohr and H(1p) of 1.5 Bohr. The k-point sampling for the semi-infinite leads is 3x3x20 and for the extended molecule is 3x3x1.

Fig. \ref{fig:au-h2} shows the mean displacement of the unconstrained atoms against the current flow, and the elongation of the H-H bond for both CP2K+SMEAGOL and SIESTA+SMEAGOL. While the exact H-H bond lengths are different, the elongation as a function of an applied bias up to 1.5 V bias is consistent. The minor differences in the absolute bond lengths can be attributed to the difference in the basis sets used for the calculations. By comparing the results to a hydrogen molecule in vacuum in the presence of an effective electric field,  Bai et al.\cite{baiCurrentinducedPhononRenormalization2016} confirmed in their work that the observed geometrical changes are are dominated by the electric current rather than by the electric field. 

We have verified that there is no significant difference in the geometry when the Green's function is evaluated using a weighted double contour (Eq. \ref{eq:density_matrix_double_contour}) instead of the single contour (Eq. \ref{eq:single_contour}) evaluation used in the original work of Bai et al.\cite{baiCurrentinducedPhononRenormalization2016}.

\subsection{Molecular dynamics of a solvated Au wire} \label{section:au_wire_solvated}

\begin{figure}[t!]
         \includegraphics[width=1\columnwidth]{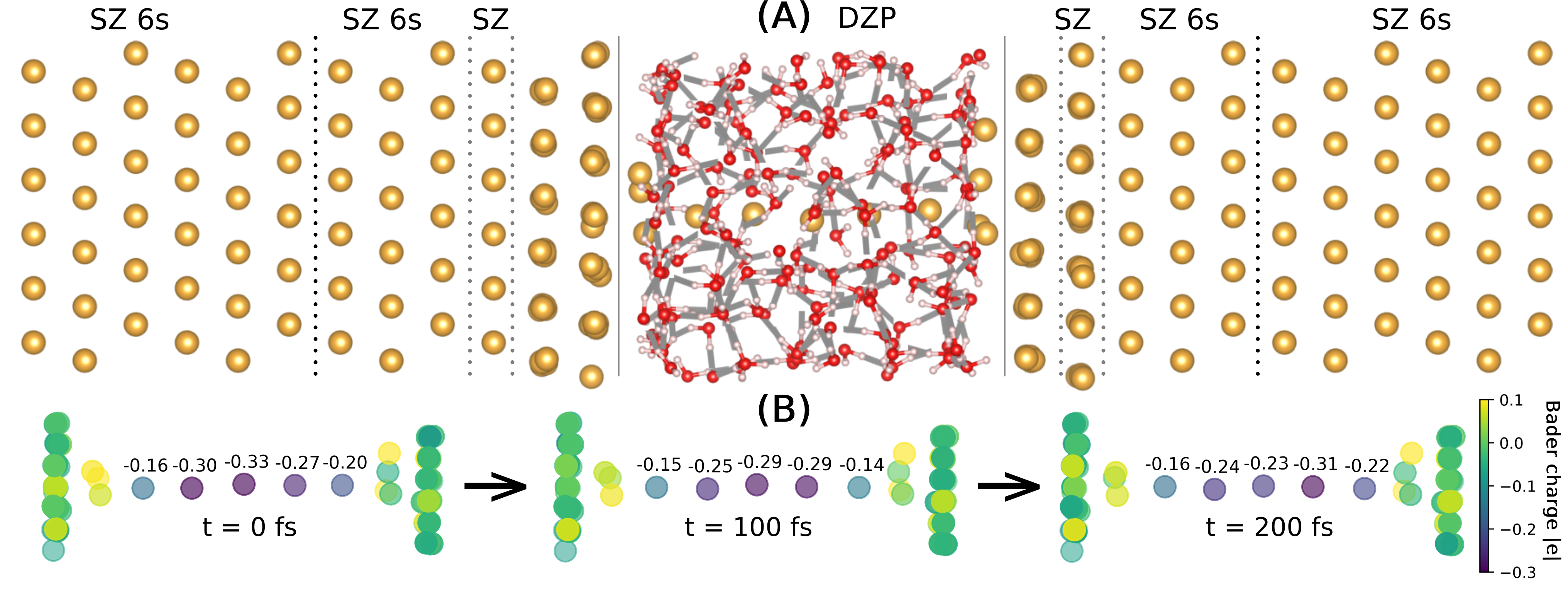}
     \caption{Molecular dynamics of a solvated Au wire using CP2K+SMEAGOL at an applied bias of 1 V. (A) DFT equilibrated starting geometry of the solvated Au wire, where yellow, white and red spheres represent Au, H and O atoms respectively. The basis set used for the Au atoms is shown above the structure, with each distinct Au region separated by a dotted black line. (B) Bader charges for the Au wire atoms, Au tip atoms and the first layer of the Au slabs calculated on the DFT equilibrated geometry as well as after 100 fs MD and after 200 fs MD.}
     \label{fig:em_au-wire_all}
\end{figure}

\begin{figure}[t!]
         \includegraphics[width=0.5\columnwidth]{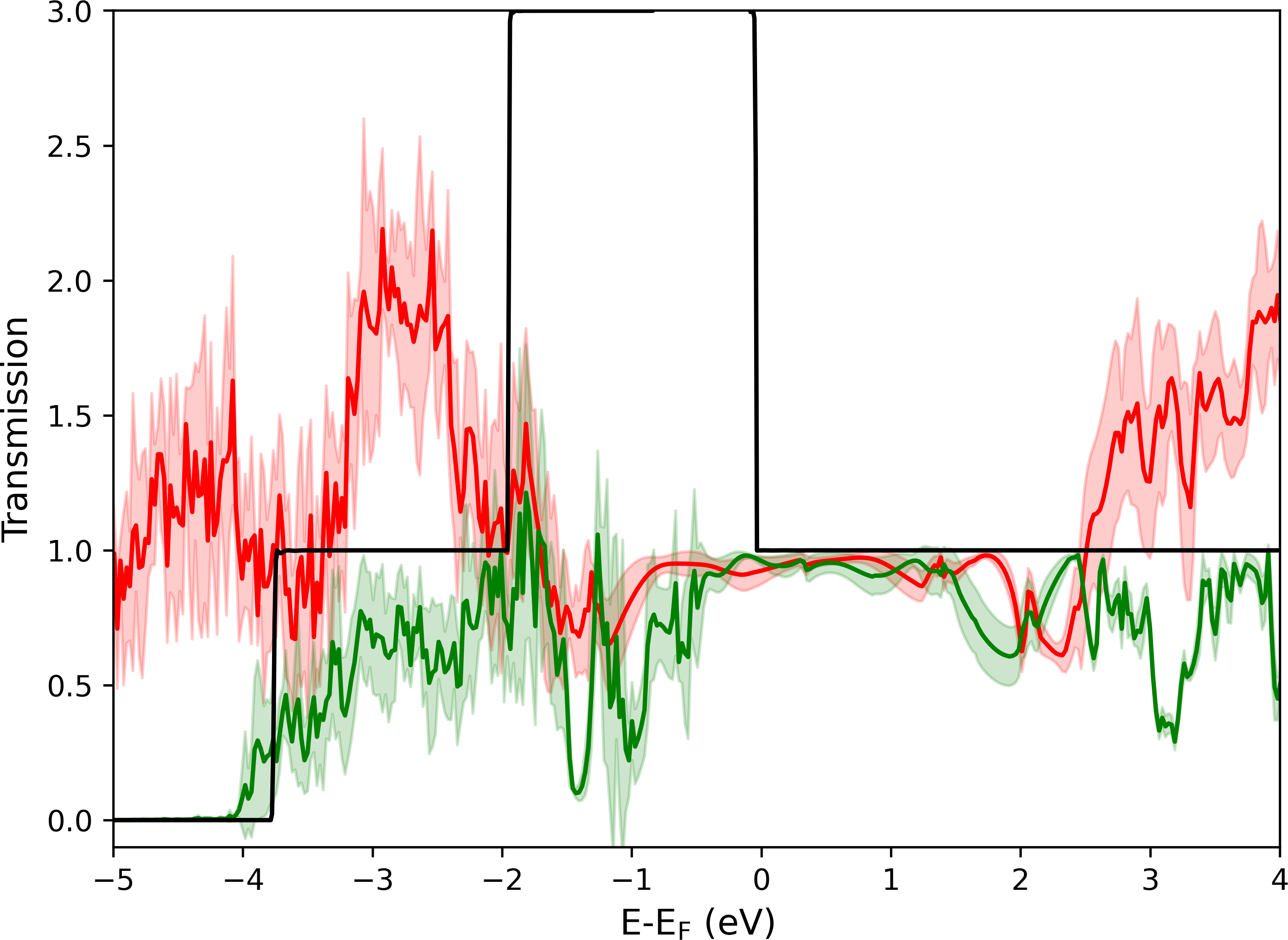}
     \caption{Average transmission calculated using CP2K+SMEAGOL at zero bias. Solvated Au wire (red) and with water molecules removed (green) sampled every 100 fs across 500 fs of DFT-MD, with comparison to the transmission for an ideal infinite Au wire in vacuum (black). The shaded region represents 1 standard deviation of the mean, shown as a solid line.}
     \label{fig:compare-transmission}
\end{figure}

Atomic size metallic contacts have been studied extensively \cite{yansonFormationManipulationMetallic1998, untiedtCalibrationLengthChain2002, sanchez-portalStiffMonatomicGold1999, torresPuzzlingStabilityMonatomic1999, sorensenMechanicalDeformationAtomicscale1998}, with interest both in their fundamental properties as well as for their possible applications in nano-electronics. An important application particularly relevant to this work is in high density memories and logic applications, e.g. in molecular switches and electrohemical metallisation memories\cite{menzelUnderstandingFilamentaryGrowth2015,menzelModelingSimulationElectrochemical2014,milanoRecentDevelopmentsPerspectives2019}, where the non equilibrium component of the force is essential to describe the change of status of the system.

Monoatomic Au wires are typically formed through experimental techniques such as scanning tunnelling microscopy (STM)\cite{binnigSurfaceStudiesScanning1982} and the mechanically controllable break junction (MCBJ)\cite{mullerConductanceSupercurrentDiscontinuities1992, wangAdvanceMechanicallyControllable2019}, however most studies are performed in vacuum and therefore neglect any solvation effects. In this work we present CP2K and CP2K+SMEAGOL MD calculations for a fully solvated monoatomic Au wire composed of a 4-atom wire, connected via two 3-atom pyramidal tips to two Au(111) slabs.

Starting from an initial structure with a straight Au wire forming Au-Au-Au bond angles of 180\textdegree, during DFT cell optimisation performed in vacuum the Au wire relaxes to form a zigzag geometry\cite{sanchez-portalStiffMonatomicGold1999} with an Au-Au-Au bond angle of around 130\textdegree. The Au wire was then fully solvated by adding 166 water molecules, equilibrated by performing 30 ns classical MD with a TIP3P water model\cite{jorgensenComparisonSimplePotential1983a} and Lennard-Jones potentials\cite{bergEvaluationOptimizationInterface2017}. The Au atoms were then allowed to relax through 9 ps of NVT DFT-MD, where the Au wire straightened and one Au wire atoms was forced into the tip geometry. The PBE functional was used, with triple-$\zeta$ plus polarisation functions for the Au(6s, 5d) electrons, O(2p, 2s) electrons and H(1s) electrons. 

To decrease the significant computational cost of subsequent CP2K+SMEAGOL calculations it is necessary to use a minimal single-$\zeta$ Au(6s) basis set for the leads atoms, which has been shown in previous work to reproduce key transport properties such as the transmission and current up to an applied bias of around 2 V\cite{odellInvestigationConductingProperties2010,toherEffectsSelfinteractionCorrections2008}. In addition the triple-$\zeta$ basis set for the solvated Au wire is reduced to a double-$\zeta$ basis set, and to avoid spurious density build-up at the interface between the different basis sets of the extended molecule and the leads additional screening regions are added. See ESI Section 1.3 for further details regarding the basis set choices and size of the leads. All Au atoms are frozen except for the 3-atom Au wire and the 7 atoms belonging to the pyramidal Au tips. With this new configuration 1.5 ps of NVT DFT-MD is performed to re-equilibrate the system, resulting in the structure shown in Fig. \ref{fig:em_au-wire_all}.

Fig. \ref{fig:compare-transmission} shows the transmission calculated for the solvated Au wire sampled across 500 fs of DFT-MD, with comparison to the transmission of an ideal infinite Au wire. Qualitatively the transmission is very similar, roughly equal to 1 between -2 eV and 2 eV. The differences in transmission between the solvated wire and the system where water was removed (red and green curves in Fig. \ref{fig:compare-transmission} ) can be ascribed to the opening of additional tunnelling channels for the electron through the junction when the water is present. More specifically, these channels become available in the same energy range where the water band arises, with the density of states shown in ESI Fig. 5.  The peak in transmission for the ideal Au wire between -2 and 0 eV is not present for the solvated Au wire, as this peak is due to Au(5d) states that are neglected in the leads atoms for the solvated Au wire. 

We also perform CP2K+SMEAGOL molecular dynamics at an applied bias of 0 V, 0.1 V and 1 V. While the total energy for such a current-carrying open system may not be conserved\cite{todorovCurrentinducedForcesSimple2014}, in practice we that the dynamics are stable with minimal long-term energy drift. See ESI Section 1.3 for further discussion of energy conservation for molecular dynamics performed for both the solvated Au wire (Section \ref{section:au_wire_solvated}) and the Au-H$_{2}$-Au junction (Section \ref{section:au_h2_au}).

Fig. \ref{fig:em_au-wire_all} shows the Bader charges of the Au wire atoms, as well as the Au tip and first layer of the Au slabs. The Bader charges are calculated as the difference in the reference charge and the atomic contribution to the electron density calculated using the algorithm developed by Henkelman et al\cite{Henkelman2006}, in units of the elementary charge |e|. With both CP2K and CP2K+SMEAGOL at zero applied bias the Au atoms belonging to the Au wire are negatively charged (-0.30, -0.32, -0.27), as well as the first atom of the pyramidal Au tips (-0.19, -0.18). The remaining Au atoms of the tips are slightly positively charged: +0.09 to the left of the Au wire, +0.05 to the right of the Au wire, while the first layer of the Au slabs are approximately neutral: -0.03 and -0.01 respectively. With an applied bias of 0.1 V there is no significant change in the Bader charges, while at an applied bias of 1 V the charge on the leftmost Au wire atom increases from -0.19 to -0.16 (+0.03), and the charge on the rightmost Au atom decreases from -0.18 to -0.20 (-0.02). Performing 200 fs CP2K+SMEAGOL MD at an applied bias of 1 V we observe electron migration effects, where the minima in the Bader charge (maxima in the local electron density) moves from the central Au wire atom (-0.30, -0.33, -0.27) to the rightmost Au wire atom (-0.24, -0.23, -0.31). This is consistent with the migration of electron density to screen the positive charge of the right electrode.

Due to the substantial cost of performing CP2K+SMEAGOL calculations only a short trajectory of 200 fs was possible, and as such work is ongoing in our group to accelerate the CP2K+SMEAGOL calculations (Section \ref{section:performance}). In the future, we hope to calculate the conductance as a function of the elongation of the Au wire and to study the breaking of the Au wire as a function of the applied bias.

\subsection{Performance} \label{section:performance}

\begin{figure}[t!]
         \includegraphics[width=0.8\columnwidth]{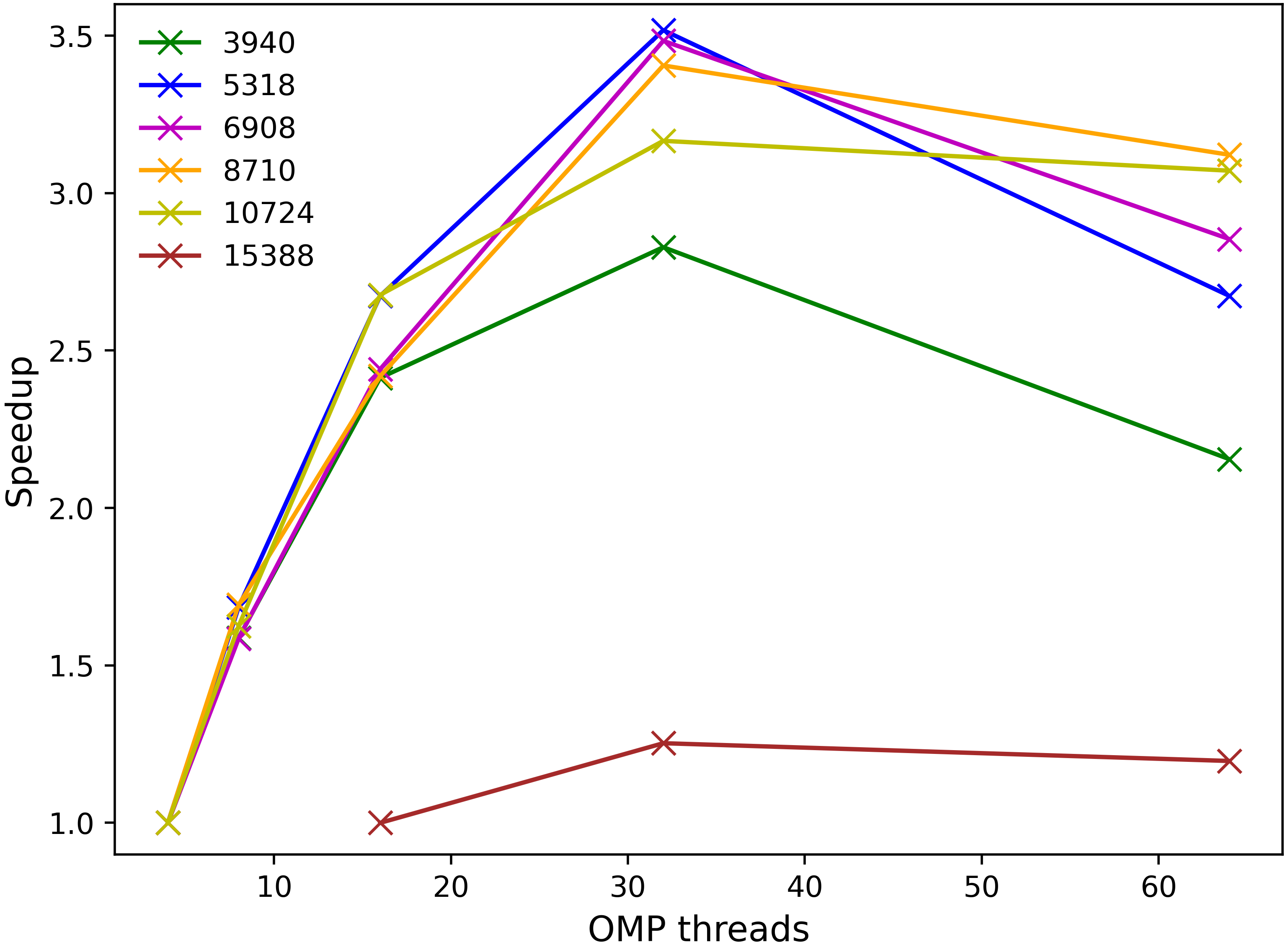}
     \caption{Speedup of CP2K+SMEAGOL as a function of the number of OpenMP threads for different system sizes. Calculations are performed for a model Au-BDT-Au junction, where the number of atoms is increased from 948 (3940 basis functions) to 3756 (15388 basis functions).}
     \label{fig:scaling}
\end{figure}

The main bottleneck in a CP2K+SMEAGOL calculation is the evaluation of the density as an integral of the Green's function (Eq. \ref{eq:total_density_matrix_left}), as many computations of the Green's function must be performed at each SCF step. In all calculations performed in this work we perform 64 computations of the Green's function along the real axis (Eq. \ref{eq:density_matrix_nonequilibrium_real}) and 32 computations of the Green's function along the complex axis (Eq. \ref{eq:density_matrix_equilibrium_complex}), such that a single diagonalisation of the Hamiltonian in a standard CP2K calculation is replaced by computing the Green's function at 96 energy points. As each computation is independent, SMEAGOL uses MPI to parallelise up to the number of real space computations multiplied by the number of kpoints. Further parallelism is available through the use of OpenMP, with both OpenMP parallelised DO loops as well as threaded LAPACK routines. The speedup of CP2K+SMEAGOL as a function of the number of OpenMP threads for different system sizes is shown in Fig. \ref{fig:scaling}, with an improvement in performance up to 32 threads for all system sizes. We find that while larger systems can be studied, CP2K+SMEAGOL performs optimally for systems with around 10'000 basis functions. We note that while SMEAGOL does include an O(N) scaling algorithm for the computation of the Green's function\cite{spellacyPerformanceAnalysisPairwise2019}, only calculating the block tridiagonal part of the inverse, we have found in practice that this increases the number of required matrix multiplications and leads to an overall decrease in performance. As such, we do not use this O(N) scaling algorithm in any calculations performed in this work.

For the solvated Au wire (Section \ref{section:au_wire_solvated}) we performed 64 computations of the real space Green's function at $\Gamma$-point, using 64 MPI processes. For each MPI process we used an additional 32 OpenMP threads, with a total of 64*32=2048 cores. All calculations were performed on the ARCHER2 UK National Supercomputing Service (AMD 2.25GHz), with an average time of 55 s per SCF step under bias in comparison to 17 s per SCF step for a corresponding CP2K calculation performed on 256 cores, a factor of 55/17=3.2 times slower on 2048/256=8 times the number of cores. These numbers are representative of the additional cost of performing a CP2K+SMEAGOL calculation, around an order of magnitude more expensive than a corresponding CP2K calculation. 

While for a typical transport calculation the additional cost of a CP2K+SMEAGOL calculation is largely insignificant, as only a small number of calculations are required to be performed under bias, for performing molecular dynamics calculations it becomes problematic. As such, we are currently investigating methods that can be used to accelerate or circumvent the need for these calculations. The use of ScaLAPACK could allow the scaling limit to be improved, using distributed memory parallelism for the matrix inversion and matrix multiplication routines. Alternatively, it could be possible to use CP2K+SMEAGOL to generate training data for a machine learning force field\cite{Behler2021,zhouMachineLearningAssisted2023}. Work towards this is ongoing in our group.

\section{Conclusion} \label{section:conclusion}

In this work we have developed a new interface between the popular DFT package CP2K and the NEGF code SMEAGOL, allowing for CP2K calculations to be performed under an applied potential and current flow. In contrast to the existing DFT-NEGF implementation available natively in CP2K, both k-point sampling and forces are available.

We have verified and benchmarked our interface against systems previously studied with SMEAGOL and the DFT code SIESTA, showing good agreement for both single point calculations and geometry re-optimisation under bias for: an infinite Au wire, a parallel-plate capacitor and a Au-H$_{2}$-Au junction. 

For the first time we are able to perform large-scale molecular dynamics simulations under realistic operating conditions using the DFT-NEGF approach. For an example of a solvated Au wire, we observe electron migration effects during a short 200 fs MD trajectory at an applied bias of 1 V. We expect that this methodology will become valuable for the emerging field of first principles electrochemistry, for example to model the effect of an applied potential or current flow through an electro-chemical cell. 

\begin{acknowledgement}

We would like to thank Lev Kantorovich for the insightful discussions that contributed to this work. We would like to thank Jad Jaafar and Dr Margherita Buraschi for performing the original molecular dynamics calculations of the solvated Au wire. This work was supported by the Engineering and Physical Sciences Research Council (grant EP/P033555/1). We would like to acknowledge the Thomas Young Centre under grant number TYC-101. We also gratefully acknowledge the use of the High-Performance Computers at Imperial College London, provided by Imperial College Research Computing Service DOI: 10.14469/hpc/2232, and the computing resources provided by STFC Scientific Computing Department’s SCARF cluster. This research also used ARCHER2 UK National Supercomputing Service (https://www.archer2.ac.uk), via our membership of the UK’s HEC Materials Chemistry Consortium, which is funded by EPSRC (EP/R029431 and EP/X035859). We are thankful for the computational support provided by Dr. Alin Marin Elena, and for the valuable support and discussions with Teodoro Laino, Marcella Iannuzzi, Ivan Runnger, and Alex Rocha, especially during the early stages of the CP2K-SMEAGOL interface implementation. 

\end{acknowledgement}

\begin{suppinfo}

The Supporting Information is available free of charge

\begin{itemize}
  \item Au wire zero bias forces, parallel-plate capacitor planer average of the charge density, more details regarding the structure of solvated Au wire system, energy conservation during CP2K+SMEAGOL MD and finite bias calculations of a Au-Melamine-Au junction.
\end{itemize}

\end{suppinfo}


\providecommand{\latin}[1]{#1}
\makeatletter
\providecommand{\doi}
  {\begingroup\let\do\@makeother\dospecials
  \catcode`\{=1 \catcode`\}=2 \doi@aux}
\providecommand{\doi@aux}[1]{\endgroup\texttt{#1}}
\makeatother
\providecommand*\mcitethebibliography{\thebibliography}
\csname @ifundefined\endcsname{endmcitethebibliography}
  {\let\endmcitethebibliography\endthebibliography}{}
\begin{mcitethebibliography}{0}
\providecommand*\natexlab[1]{#1}
\providecommand*\mciteSetBstSublistMode[1]{}
\providecommand*\mciteSetBstMaxWidthForm[2]{}
\providecommand*\mciteBstWouldAddEndPuncttrue
  {\def\EndOfBibitem{\unskip.}}
\providecommand*\mciteBstWouldAddEndPunctfalse
  {\let\EndOfBibitem\relax}
\providecommand*\mciteSetBstMidEndSepPunct[3]{}
\providecommand*\mciteSetBstSublistLabelBeginEnd[3]{}
\providecommand*\EndOfBibitem{}
\mciteSetBstSublistMode{f}
\mciteSetBstMaxWidthForm{subitem}{(\alph{mcitesubitemcount})}
\mciteSetBstSublistLabelBeginEnd
  {\mcitemaxwidthsubitemform\space}
  {\relax}
  {\relax}

\end{mcitethebibliography}


\begin{mcitethebibliography}{59}
\providecommand*\natexlab[1]{#1}
\providecommand*\mciteSetBstSublistMode[1]{}
\providecommand*\mciteSetBstMaxWidthForm[2]{}
\providecommand*\mciteBstWouldAddEndPuncttrue
  {\def\EndOfBibitem{\unskip.}}
\providecommand*\mciteBstWouldAddEndPunctfalse
  {\let\EndOfBibitem\relax}
\providecommand*\mciteSetBstMidEndSepPunct[3]{}
\providecommand*\mciteSetBstSublistLabelBeginEnd[3]{}
\providecommand*\EndOfBibitem{}
\mciteSetBstSublistMode{f}
\mciteSetBstMaxWidthForm{subitem}{(\alph{mcitesubitemcount})}
\mciteSetBstSublistLabelBeginEnd
  {\mcitemaxwidthsubitemform\space}
  {\relax}
  {\relax}

\bibitem[Nielsen \latin{et~al.}(2015)Nielsen, Bj{\"o}rketun, Hansen, and
  Rossmeisl]{nielsenFirstPrinciplesModeling2015}
Nielsen,~M.; Bj{\"o}rketun,~M.~E.; Hansen,~M.~H.; Rossmeisl,~J. Towards First
  Principles Modeling of Electrochemical Electrode--Electrolyte Interfaces.
  \emph{Surface Science} \textbf{2015}, \emph{631}, 2--7\relax
\mciteBstWouldAddEndPuncttrue
\mciteSetBstMidEndSepPunct{\mcitedefaultmidpunct}
{\mcitedefaultendpunct}{\mcitedefaultseppunct}\relax
\EndOfBibitem
\bibitem[Sundararaman \latin{et~al.}(2022)Sundararaman, {Vigil-Fowler}, and
  Schwarz]{sundararamanImprovingAccuracyAtomistic2022}
Sundararaman,~R.; {Vigil-Fowler},~D.; Schwarz,~K. Improving the {{Accuracy}} of
  {{Atomistic Simulations}} of the {{Electrochemical Interface}}. \emph{Chem.
  Rev.} \textbf{2022}, \emph{122}, 10651--10674\relax
\mciteBstWouldAddEndPuncttrue
\mciteSetBstMidEndSepPunct{\mcitedefaultmidpunct}
{\mcitedefaultendpunct}{\mcitedefaultseppunct}\relax
\EndOfBibitem
\bibitem[Khatib \latin{et~al.}(2021)Khatib, Kumar, Sanvito, Sulpizi, and
  Cucinotta]{Khatib2021}
Khatib,~R.; Kumar,~A.; Sanvito,~S.; Sulpizi,~M.; Cucinotta,~C.~S.
  Electrochimica {{Acta The}} Nanoscale Structure of the {{Pt-water}} Double
  Layer under Bias Revealed. \emph{Electrochimica Acta} \textbf{2021},
  \emph{391}, 138875\relax
\mciteBstWouldAddEndPuncttrue
\mciteSetBstMidEndSepPunct{\mcitedefaultmidpunct}
{\mcitedefaultendpunct}{\mcitedefaultseppunct}\relax
\EndOfBibitem
\bibitem[Xue \latin{et~al.}(2002)Xue, Datta, and
  Ratner]{xueFirstprinciplesBasedMatrix2002}
Xue,~Y.; Datta,~S.; Ratner,~M.~A. First-Principles Based Matrix {{Green}}'s
  Function Approach to Molecular Electronic Devices: General Formalism.
  \emph{Chemical Physics} \textbf{2002}, \emph{281}, 151--170\relax
\mciteBstWouldAddEndPuncttrue
\mciteSetBstMidEndSepPunct{\mcitedefaultmidpunct}
{\mcitedefaultendpunct}{\mcitedefaultseppunct}\relax
\EndOfBibitem
\bibitem[Taylor \latin{et~al.}(2001)Taylor, Guo, and
  Wang]{taylorInitioModelingQuantum2001}
Taylor,~J.; Guo,~H.; Wang,~J. {\emph{Ab Initio}} Modeling of Quantum Transport
  Properties of Molecular Electronic Devices. \emph{Phys. Rev. B}
  \textbf{2001}, \emph{63}, 245407\relax
\mciteBstWouldAddEndPuncttrue
\mciteSetBstMidEndSepPunct{\mcitedefaultmidpunct}
{\mcitedefaultendpunct}{\mcitedefaultseppunct}\relax
\EndOfBibitem
\bibitem[Evers \latin{et~al.}(2020)Evers, Koryt\'ar, Tewari, and van
  Ruitenbeek]{RevModPhys.92.035001}
Evers,~F.; Koryt\'ar,~R.; Tewari,~S.; van Ruitenbeek,~J.~M. Advances and
  challenges in single-molecule electron transport. \emph{Rev. Mod. Phys.}
  \textbf{2020}, \emph{92}, 035001\relax
\mciteBstWouldAddEndPuncttrue
\mciteSetBstMidEndSepPunct{\mcitedefaultmidpunct}
{\mcitedefaultendpunct}{\mcitedefaultseppunct}\relax
\EndOfBibitem
\bibitem[Sanvito(2011)]{10.1039/BK9781849731331-00179}
Sanvito,~S. \emph{{Computational Nanoscience}}; The Royal Society of Chemistry,
  2011\relax
\mciteBstWouldAddEndPuncttrue
\mciteSetBstMidEndSepPunct{\mcitedefaultmidpunct}
{\mcitedefaultendpunct}{\mcitedefaultseppunct}\relax
\EndOfBibitem
\bibitem[Rocha \latin{et~al.}(2006)Rocha, {Garc{\'i}a-Su{\'a}rez}, Bailey,
  Lambert, Ferrer, and Sanvito]{rochaSpinMolecularElectronics2006}
Rocha,~A.~R.; {Garc{\'i}a-Su{\'a}rez},~V.~M.; Bailey,~S.; Lambert,~C.;
  Ferrer,~J.; Sanvito,~S. Spin and Molecular Electronics in Atomically
  Generated Orbital Landscapes. \emph{Phys. Rev. B} \textbf{2006}, \emph{73},
  085414\relax
\mciteBstWouldAddEndPuncttrue
\mciteSetBstMidEndSepPunct{\mcitedefaultmidpunct}
{\mcitedefaultendpunct}{\mcitedefaultseppunct}\relax
\EndOfBibitem
\bibitem[Pomorski \latin{et~al.}(2004)Pomorski, Pastewka, Roland, Guo, and
  Wang]{pomorskiCapacitanceInducedCharges2004}
Pomorski,~P.; Pastewka,~L.; Roland,~C.; Guo,~H.; Wang,~J. Capacitance, Induced
  Charges, and Bound States of Biased Carbon Nanotube Systems. \emph{Phys. Rev.
  B} \textbf{2004}, \emph{69}, 115418\relax
\mciteBstWouldAddEndPuncttrue
\mciteSetBstMidEndSepPunct{\mcitedefaultmidpunct}
{\mcitedefaultendpunct}{\mcitedefaultseppunct}\relax
\EndOfBibitem
\bibitem[Chulkov \latin{et~al.}()Chulkov, Watkins, Kantorovich, and
  Bethune]{chulkovCP2KElectronTransport}
Chulkov,~S.~K.; Watkins,~M.~B.; Kantorovich,~L.~N.; Bethune,~I. {{CP2K}} ---
  {{Electron Transport}} Based on {{Non-Equilibrium-Green}}'s-{{Functions
  Method}}: {{eCSE}} 08-09 {{Technical Report}}. 11\relax
\mciteBstWouldAddEndPuncttrue
\mciteSetBstMidEndSepPunct{\mcitedefaultmidpunct}
{\mcitedefaultendpunct}{\mcitedefaultseppunct}\relax
\EndOfBibitem
\bibitem[Smidstrup \latin{et~al.}(2020)Smidstrup, Markussen, Vancraeyveld,
  Wellendorff, Schneider, Gunst, Verstichel, Stradi, Khomyakov, {Vej-Hansen},
  Lee, Chill, Rasmussen, Penazzi, Corsetti, Ojanper{\"a}, Jensen, Palsgaard,
  Martinez, Blom, Brandbyge, and
  Stokbro]{smidstrupQuantumATKIntegratedPlatform2020}
Smidstrup,~S.; Markussen,~T.; Vancraeyveld,~P.; Wellendorff,~J.; Schneider,~J.;
  Gunst,~T.; Verstichel,~B.; Stradi,~D.; Khomyakov,~P.~A.; {Vej-Hansen},~U.~G.;
  Lee,~M.-E.; Chill,~S.~T.; Rasmussen,~F.; Penazzi,~G.; Corsetti,~F.;
  Ojanper{\"a},~A.; Jensen,~K.; Palsgaard,~M. L.~N.; Martinez,~U.; Blom,~A.;
  Brandbyge,~M.; Stokbro,~K. {{QuantumATK}}: An Integrated Platform of
  Electronic and Atomic-Scale Modelling Tools. \emph{J. Phys.: Condens. Matter}
  \textbf{2020}, \emph{32}, 015901\relax
\mciteBstWouldAddEndPuncttrue
\mciteSetBstMidEndSepPunct{\mcitedefaultmidpunct}
{\mcitedefaultendpunct}{\mcitedefaultseppunct}\relax
\EndOfBibitem
\bibitem[Papior \latin{et~al.}(2017)Papior, Lorente, Frederiksen, Garc{\'i}a,
  and Brandbyge]{papiorImprovementsNonequilibriumTransport2017}
Papior,~N.; Lorente,~N.; Frederiksen,~T.; Garc{\'i}a,~A.; Brandbyge,~M.
  Improvements on Non-Equilibrium and Transport {{Green}} Function Techniques:
  {{The}} next-Generation Transiesta. \emph{Computer Physics Communications}
  \textbf{2017}, \emph{212}, 8--24\relax
\mciteBstWouldAddEndPuncttrue
\mciteSetBstMidEndSepPunct{\mcitedefaultmidpunct}
{\mcitedefaultendpunct}{\mcitedefaultseppunct}\relax
\EndOfBibitem
\bibitem[Bagrets(2013)]{bagretsSpinPolarizedElectronTransport2013}
Bagrets,~A. Spin-{{Polarized Electron Transport Across Metal}}--{{Organic
  Molecules}}: {{A Density Functional Theory Approach}}. \emph{J. Chem. Theory
  Comput.} \textbf{2013}, \emph{9}, 2801--2815\relax
\mciteBstWouldAddEndPuncttrue
\mciteSetBstMidEndSepPunct{\mcitedefaultmidpunct}
{\mcitedefaultendpunct}{\mcitedefaultseppunct}\relax
\EndOfBibitem
\bibitem[Chen \latin{et~al.}(2012)Chen, Thygesen, and
  Jacobsen]{chenInitioNonequilibriumQuantum2012}
Chen,~J.; Thygesen,~K.~S.; Jacobsen,~K.~W. {\emph{Ab Initio}} Nonequilibrium
  Quantum Transport and Forces with the Real-Space Projector Augmented Wave
  Method. \emph{Phys. Rev. B} \textbf{2012}, \emph{85}, 155140\relax
\mciteBstWouldAddEndPuncttrue
\mciteSetBstMidEndSepPunct{\mcitedefaultmidpunct}
{\mcitedefaultendpunct}{\mcitedefaultseppunct}\relax
\EndOfBibitem
\bibitem[Ozaki \latin{et~al.}(2010)Ozaki, Nishio, and
  Kino]{ozakiEfficientImplementationNonequilibrium2010}
Ozaki,~T.; Nishio,~K.; Kino,~H. Efficient Implementation of the Nonequilibrium
  {{Green}} Function Method for Electronic Transport Calculations. \emph{Phys.
  Rev. B} \textbf{2010}, \emph{81}, 035116\relax
\mciteBstWouldAddEndPuncttrue
\mciteSetBstMidEndSepPunct{\mcitedefaultmidpunct}
{\mcitedefaultendpunct}{\mcitedefaultseppunct}\relax
\EndOfBibitem
\bibitem[Saha \latin{et~al.}(2009)Saha, Lu, Bernholc, and
  Meunier]{sahaFirstprinciplesMethodologyQuantum2009}
Saha,~K.~K.; Lu,~W.; Bernholc,~J.; Meunier,~V. First-Principles Methodology for
  Quantum Transport in Multiterminal Junctions. \emph{The Journal of Chemical
  Physics} \textbf{2009}, \emph{131}, 164105\relax
\mciteBstWouldAddEndPuncttrue
\mciteSetBstMidEndSepPunct{\mcitedefaultmidpunct}
{\mcitedefaultendpunct}{\mcitedefaultseppunct}\relax
\EndOfBibitem
\bibitem[Kami{\'n}ski \latin{et~al.}(2016)Kami{\'n}ski, Topolnicki, Hapala,
  Jel{\'i}nek, and Kucharczyk]{kaminskiTuningConductanceBenzenebased2016}
Kami{\'n}ski,~W.; Topolnicki,~R.; Hapala,~P.; Jel{\'i}nek,~P.; Kucharczyk,~R.
  Tuning the Conductance of Benzene-Based Single-Molecule Junctions.
  \emph{Organic Electronics} \textbf{2016}, \emph{34}, 254--261\relax
\mciteBstWouldAddEndPuncttrue
\mciteSetBstMidEndSepPunct{\mcitedefaultmidpunct}
{\mcitedefaultendpunct}{\mcitedefaultseppunct}\relax
\EndOfBibitem
\bibitem[Pedroza \latin{et~al.}(2018)Pedroza, Brandimarte, Rocha, and
  {Fern{\'a}ndez-Serra}]{pedrozaBiasdependentLocalStructure2018}
Pedroza,~L.~S.; Brandimarte,~P.; Rocha,~A.~R.; {Fern{\'a}ndez-Serra},~M.-V.
  Bias-Dependent Local Structure of Water Molecules at a Metallic Interface.
  \emph{Chem. Sci.} \textbf{2018}, \emph{9}, 62--69\relax
\mciteBstWouldAddEndPuncttrue
\mciteSetBstMidEndSepPunct{\mcitedefaultmidpunct}
{\mcitedefaultendpunct}{\mcitedefaultseppunct}\relax
\EndOfBibitem
\bibitem[Choi \latin{et~al.}(2023)Choi, Johnson, Fomina, Darvish, Lang, Shin,
  Kim, and Jang]{choiElectronTransportNanoconfined2023}
Choi,~J.~I.; Johnson,~C.; Fomina,~N.; Darvish,~A.; Lang,~C.; Shin,~Y.~S.;
  Kim,~H.~S.; Jang,~S.~S. Electron {{Transport}} through {{Nanoconfined
  Ferrocene Solution}}: {{Density Functional Theory}}-{{Nonequilibrium Green
  Function Approach}}. \emph{J. Phys. Chem. C} \textbf{2023}, \emph{127},
  2666--2674\relax
\mciteBstWouldAddEndPuncttrue
\mciteSetBstMidEndSepPunct{\mcitedefaultmidpunct}
{\mcitedefaultendpunct}{\mcitedefaultseppunct}\relax
\EndOfBibitem
\bibitem[K{\"u}hne \latin{et~al.}(2020)K{\"u}hne, Iannuzzi, Del~Ben, Rybkin,
  Seewald, Stein, Laino, Khaliullin, Sch{\"u}tt, Schiffmann, Golze, Wilhelm,
  Chulkov, {Bani-Hashemian}, Weber, Bor{\v s}tnik, Taillefumier, Jakobovits,
  Lazzaro, Pabst, M{\"u}ller, Schade, Guidon, Andermatt, Holmberg, Schenter,
  Hehn, Bussy, Belleflamme, Tabacchi, Gl{\"o}{\ss}, Lass, Bethune, Mundy,
  Plessl, Watkins, VandeVondele, Krack, Hutter, Borstnik, Taillefumier,
  Jakobovits, Lazzaro, Pabst, M{\"u}ller, Schade, Guidon, Andermatt, Holmberg,
  Schenter, Hehn, Bussy, Belleflamme, Tabacchi, Gl{\"o}{\ss}, Lass, Bethune,
  Mundy, Plessl, Watkins, VandeVondele, Krack, and Hutter]{Kuhne2020}
K{\"u}hne,~T.~D.; Iannuzzi,~M.; Del~Ben,~M.; Rybkin,~V.~V.; Seewald,~P.;
  Stein,~F.; Laino,~T.; Khaliullin,~R.~Z.; Sch{\"u}tt,~O.; Schiffmann,~F.;
  Golze,~D.; Wilhelm,~J.; Chulkov,~S.; {Bani-Hashemian},~M.~H.; Weber,~V.;
  Bor{\v s}tnik,~U.; Taillefumier,~M.; Jakobovits,~A.~S.; Lazzaro,~A.;
  Pabst,~H.; M{\"u}ller,~T.; Schade,~R.; Guidon,~M.; Andermatt,~S.;
  Holmberg,~N.; Schenter,~G.~K.; Hehn,~A.; Bussy,~A.; Belleflamme,~F.;
  Tabacchi,~G.; Gl{\"o}{\ss},~A.; Lass,~M.; Bethune,~I.; Mundy,~C.~J.;
  Plessl,~C.; Watkins,~M.; VandeVondele,~J.; Krack,~M.; Hutter,~J.;
  Borstnik,~U.; Taillefumier,~M.; Jakobovits,~A.~S.; Lazzaro,~A.; Pabst,~H.;
  M{\"u}ller,~T.; Schade,~R.; Guidon,~M.; Andermatt,~S.; Holmberg,~N.;
  Schenter,~G.~K.; Hehn,~A.; Bussy,~A.; Belleflamme,~F.; Tabacchi,~G.;
  Gl{\"o}{\ss},~A.; Lass,~M.; Bethune,~I.; Mundy,~C.~J.; Plessl,~C.;
  Watkins,~M.; VandeVondele,~J.; Krack,~M.; Hutter,~J. {{CP2K}}: {{An
  Electronic Structure}} and {{Molecular Dynamics Software Package}} --
  {{Quickstep}}: {{Efficient}} and {{Accurate Electronic Structure
  Calculations}}. \emph{The Journal of Chemical Physics} \textbf{2020},
  \emph{194103}, 194103\relax
\mciteBstWouldAddEndPuncttrue
\mciteSetBstMidEndSepPunct{\mcitedefaultmidpunct}
{\mcitedefaultendpunct}{\mcitedefaultseppunct}\relax
\EndOfBibitem
\bibitem[Soler \latin{et~al.}(2002)Soler, Artacho, Gale, Garc{\'i}a, Junquera,
  Ordej{\'o}n, and {S{\'a}nchez-Portal}]{solerSIESTAMethodInitio2002}
Soler,~J.~M.; Artacho,~E.; Gale,~J.~D.; Garc{\'i}a,~A.; Junquera,~J.;
  Ordej{\'o}n,~P.; {S{\'a}nchez-Portal},~D. The {{SIESTA}} Method for {\emph{Ab
  Initio}} Order- {{{\emph{N}}}} Materials Simulation. \emph{J. Phys.: Condens.
  Matter} \textbf{2002}, \emph{14}, 2745--2779\relax
\mciteBstWouldAddEndPuncttrue
\mciteSetBstMidEndSepPunct{\mcitedefaultmidpunct}
{\mcitedefaultendpunct}{\mcitedefaultseppunct}\relax
\EndOfBibitem
\bibitem[Garc{\'i}a \latin{et~al.}(2020)Garc{\'i}a, Papior, Akhtar, Artacho,
  Blum, Bosoni, Brandimarte, Brandbyge, Cerd{\'a}, Corsetti, Cuadrado, Dikan,
  Ferrer, Gale, {Garc{\'i}a-Fern{\'a}ndez}, {Garc{\'i}a-Su{\'a}rez},
  Garc{\'i}a, Huhs, Illera, Koryt{\'a}r, Koval, Lebedeva, Lin,
  {L{\'o}pez-Tarifa}, Mayo, Mohr, Ordej{\'o}n, Postnikov, Pouillon, Pruneda,
  Robles, {S{\'a}nchez-Portal}, Soler, Ullah, Yu, and
  Junquera]{garciaSIESTARecentDevelopments2020}
Garc{\'i}a,~A.; Papior,~N.; Akhtar,~A.; Artacho,~E.; Blum,~V.; Bosoni,~E.;
  Brandimarte,~P.; Brandbyge,~M.; Cerd{\'a},~J.~I.; Corsetti,~F.; Cuadrado,~R.;
  Dikan,~V.; Ferrer,~J.; Gale,~J.; {Garc{\'i}a-Fern{\'a}ndez},~P.;
  {Garc{\'i}a-Su{\'a}rez},~V.~M.; Garc{\'i}a,~S.; Huhs,~G.; Illera,~S.;
  Koryt{\'a}r,~R.; Koval,~P.; Lebedeva,~I.; Lin,~L.; {L{\'o}pez-Tarifa},~P.;
  Mayo,~S.~G.; Mohr,~S.; Ordej{\'o}n,~P.; Postnikov,~A.; Pouillon,~Y.;
  Pruneda,~M.; Robles,~R.; {S{\'a}nchez-Portal},~D.; Soler,~J.~M.; Ullah,~R.;
  Yu,~V. W.-z.; Junquera,~J. {{SIESTA}}: {{Recent}} Developments and
  Applications: {{Recent}} Developments and Applications. \emph{The Journal of
  Chemical Physics} \textbf{2020}, \emph{152}, 204108\relax
\mciteBstWouldAddEndPuncttrue
\mciteSetBstMidEndSepPunct{\mcitedefaultmidpunct}
{\mcitedefaultendpunct}{\mcitedefaultseppunct}\relax
\EndOfBibitem
\bibitem[Calderara \latin{et~al.}(2015)Calderara, Br{\"u}ck, Pedersen,
  {Bani-Hashemian}, VandeVondele, and Luisier]{calderaraPushingBackLimit2015}
Calderara,~M.; Br{\"u}ck,~S.; Pedersen,~A.; {Bani-Hashemian},~M.~H.;
  VandeVondele,~J.; Luisier,~M. Pushing Back the Limit of {\emph{Ab-Initio}}
  Quantum Transport Simulations on Hybrid Supercomputers. Proceedings of the
  {{International Conference}} for {{High Performance Computing}},
  {{Networking}}, {{Storage}} and {{Analysis}}. Austin Texas, 2015; pp
  1--12\relax
\mciteBstWouldAddEndPuncttrue
\mciteSetBstMidEndSepPunct{\mcitedefaultmidpunct}
{\mcitedefaultendpunct}{\mcitedefaultseppunct}\relax
\EndOfBibitem
\bibitem[Brandbyge \latin{et~al.}(2002)Brandbyge, Mozos, Ordej{\'o}n, Taylor,
  and Stokbro]{brandbygeDensityfunctionalMethodNonequilibrium2002}
Brandbyge,~M.; Mozos,~J.-L.; Ordej{\'o}n,~P.; Taylor,~J.; Stokbro,~K.
  Density-Functional Method for Nonequilibrium Electron Transport. \emph{Phys.
  Rev. B} \textbf{2002}, \emph{65}, 165401\relax
\mciteBstWouldAddEndPuncttrue
\mciteSetBstMidEndSepPunct{\mcitedefaultmidpunct}
{\mcitedefaultendpunct}{\mcitedefaultseppunct}\relax
\EndOfBibitem
\bibitem[B{\"u}ttiker \latin{et~al.}(1985)B{\"u}ttiker, Imry, Landauer, and
  Pinhas]{buttikerGeneralizedManychannelConductance1985}
B{\"u}ttiker,~M.; Imry,~Y.; Landauer,~R.; Pinhas,~S. Generalized Many-Channel
  Conductance Formula with Application to Small Rings. \emph{Phys. Rev. B}
  \textbf{1985}, \emph{31}, 6207--6215\relax
\mciteBstWouldAddEndPuncttrue
\mciteSetBstMidEndSepPunct{\mcitedefaultmidpunct}
{\mcitedefaultendpunct}{\mcitedefaultseppunct}\relax
\EndOfBibitem
\bibitem[Meir and Wingreen(1992)Meir, and
  Wingreen]{meirLandauerFormulaCurrent1992}
Meir,~Y.; Wingreen,~N.~S. Landauer Formula for the Current through an
  Interacting Electron Region. \emph{Phys. Rev. Lett.} \textbf{1992},
  \emph{68}, 2512--2515\relax
\mciteBstWouldAddEndPuncttrue
\mciteSetBstMidEndSepPunct{\mcitedefaultmidpunct}
{\mcitedefaultendpunct}{\mcitedefaultseppunct}\relax
\EndOfBibitem
\bibitem[Todorov \latin{et~al.}(2014)Todorov, Dundas, L{\"u}, Brandbyge, and
  Hedeg{\aa}rd]{todorovCurrentinducedForcesSimple2014}
Todorov,~T.~N.; Dundas,~D.; L{\"u},~J.-T.; Brandbyge,~M.; Hedeg{\aa}rd,~P.
  Current-Induced Forces: A Simple Derivation. \emph{Eur. J. Phys.}
  \textbf{2014}, \emph{35}, 065004\relax
\mciteBstWouldAddEndPuncttrue
\mciteSetBstMidEndSepPunct{\mcitedefaultmidpunct}
{\mcitedefaultendpunct}{\mcitedefaultseppunct}\relax
\EndOfBibitem
\bibitem[Zhang \latin{et~al.}(2011)Zhang, Rungger, Sanvito, and
  Hou]{zhangCurrentinducedEnergyBarrier2011}
Zhang,~R.; Rungger,~I.; Sanvito,~S.; Hou,~S. Current-Induced Energy Barrier
  Suppression for Electromigration from First Principles. \emph{Phys. Rev. B}
  \textbf{2011}, \emph{84}, 085445\relax
\mciteBstWouldAddEndPuncttrue
\mciteSetBstMidEndSepPunct{\mcitedefaultmidpunct}
{\mcitedefaultendpunct}{\mcitedefaultseppunct}\relax
\EndOfBibitem
\bibitem[Di~Ventra and Pantelides(2000)Di~Ventra, and
  Pantelides]{diventraHellmannFeynmanTheoremDefinition2000}
Di~Ventra,~M.; Pantelides,~S.~T. Hellmann-{{Feynman}} Theorem and the
  Definition of Forces in Quantum Time-Dependent and Transport Problems.
  \emph{Phys. Rev. B} \textbf{2000}, \emph{61}, 16207--16212\relax
\mciteBstWouldAddEndPuncttrue
\mciteSetBstMidEndSepPunct{\mcitedefaultmidpunct}
{\mcitedefaultendpunct}{\mcitedefaultseppunct}\relax
\EndOfBibitem
\bibitem[Pulay(1969)]{pulayInitioCalculationForce1969}
Pulay,~P. {\emph{Ab Initio}} Calculation of Force Constants and Equilibrium
  Geometries in Polyatomic Molecules: {{I}}. {{Theory}}. \emph{Molecular
  Physics} \textbf{1969}, \emph{17}, 197--204\relax
\mciteBstWouldAddEndPuncttrue
\mciteSetBstMidEndSepPunct{\mcitedefaultmidpunct}
{\mcitedefaultendpunct}{\mcitedefaultseppunct}\relax
\EndOfBibitem
\bibitem[Vandevondele \latin{et~al.}(2005)Vandevondele, Krack, Mohamed,
  Parrinello, Chassaing, and Hutter]{Vandevondele2005}
Vandevondele,~J.; Krack,~M.; Mohamed,~F.; Parrinello,~M.; Chassaing,~T.;
  Hutter,~J. Quickstep: {{Fast}} and Accurate Density Functional Calculations
  Using a Mixed {{Gaussian}} and Plane Waves Approach. \emph{Computer Physics
  Communications} \textbf{2005}, \emph{167}, 103--128\relax
\mciteBstWouldAddEndPuncttrue
\mciteSetBstMidEndSepPunct{\mcitedefaultmidpunct}
{\mcitedefaultendpunct}{\mcitedefaultseppunct}\relax
\EndOfBibitem
\bibitem[Pedroza \latin{et~al.}(2018)Pedroza, Brandimarte, Rocha, and
  {Fern{\'a}ndez-Serra}]{pedrozaBiasdependentLocalStructure2018a}
Pedroza,~L.~S.; Brandimarte,~P.; Rocha,~A.~R.; {Fern{\'a}ndez-Serra},~M.-V.
  Bias-Dependent Local Structure of Water Molecules at a Metallic Interface.
  \emph{Chem. Sci.} \textbf{2018}, \emph{9}, 62--69\relax
\mciteBstWouldAddEndPuncttrue
\mciteSetBstMidEndSepPunct{\mcitedefaultmidpunct}
{\mcitedefaultendpunct}{\mcitedefaultseppunct}\relax
\EndOfBibitem
\bibitem[Dundas \latin{et~al.}(2009)Dundas, McEniry, and
  Todorov]{dundasCurrentdrivenAtomicWaterwheels2009}
Dundas,~D.; McEniry,~E.~J.; Todorov,~T.~N. Current-Driven Atomic Waterwheels.
  \emph{Nature Nanotech} \textbf{2009}, \emph{4}, 99--102\relax
\mciteBstWouldAddEndPuncttrue
\mciteSetBstMidEndSepPunct{\mcitedefaultmidpunct}
{\mcitedefaultendpunct}{\mcitedefaultseppunct}\relax
\EndOfBibitem
\bibitem[Csonka \latin{et~al.}(2006)Csonka, Halbritter, and
  Mih{\'a}ly]{csonkaPullingGoldNanowires2006}
Csonka,~{\relax Sz}.; Halbritter,~A.; Mih{\'a}ly,~G. Pulling Gold Nanowires
  with a Hydrogen Clamp: {{Strong}} Interactions of Hydrogen Molecules with
  Gold Nanojunctions. \emph{Phys. Rev. B} \textbf{2006}, \emph{73},
  075405\relax
\mciteBstWouldAddEndPuncttrue
\mciteSetBstMidEndSepPunct{\mcitedefaultmidpunct}
{\mcitedefaultendpunct}{\mcitedefaultseppunct}\relax
\EndOfBibitem
\bibitem[Csonka \latin{et~al.}(2003)Csonka, Halbritter, Mih{\'a}ly, Jurdik,
  Shklyarevskii, Speller, and
  Van~Kempen]{csonkaFractionalConductanceHydrogenEmbedded2003}
Csonka,~{\relax Sz}.; Halbritter,~A.; Mih{\'a}ly,~G.; Jurdik,~E.;
  Shklyarevskii,~O.~I.; Speller,~S.; Van~Kempen,~H. Fractional {{Conductance}}
  in {{Hydrogen-Embedded Gold Nanowires}}. \emph{Phys. Rev. Lett.}
  \textbf{2003}, \emph{90}, 116803\relax
\mciteBstWouldAddEndPuncttrue
\mciteSetBstMidEndSepPunct{\mcitedefaultmidpunct}
{\mcitedefaultendpunct}{\mcitedefaultseppunct}\relax
\EndOfBibitem
\bibitem[Barnett \latin{et~al.}(2004)Barnett, H{\"a}kkinen, Scherbakov, and
  Landman]{barnettHydrogenWeldingHydrogen2004}
Barnett,~R.~N.; H{\"a}kkinen,~H.; Scherbakov,~A.~G.; Landman,~U. Hydrogen
  {{Welding}} and {{Hydrogen Switches}} in a {{Monatomic Gold Nanowire}}.
  \emph{Nano Lett.} \textbf{2004}, \emph{4}, 1845--1852\relax
\mciteBstWouldAddEndPuncttrue
\mciteSetBstMidEndSepPunct{\mcitedefaultmidpunct}
{\mcitedefaultendpunct}{\mcitedefaultseppunct}\relax
\EndOfBibitem
\bibitem[Jel{\'i}nek \latin{et~al.}(2006)Jel{\'i}nek, P{\'e}rez, Ortega, and
  Flores]{jelinekHydrogenDissociationAu2006}
Jel{\'i}nek,~P.; P{\'e}rez,~R.; Ortega,~J.; Flores,~F. Hydrogen
  {{Dissociation}} over {{Au Nanowires}} and the {{Fractional Conductance
  Quantum}}. \emph{Phys. Rev. Lett.} \textbf{2006}, \emph{96}, 046803\relax
\mciteBstWouldAddEndPuncttrue
\mciteSetBstMidEndSepPunct{\mcitedefaultmidpunct}
{\mcitedefaultendpunct}{\mcitedefaultseppunct}\relax
\EndOfBibitem
\bibitem[Bai \latin{et~al.}(2016)Bai, Cucinotta, Jiang, Wang, Wang, Rungger,
  Sanvito, and Hou]{baiCurrentinducedPhononRenormalization2016}
Bai,~M.; Cucinotta,~C.~S.; Jiang,~Z.; Wang,~H.; Wang,~Y.; Rungger,~I.;
  Sanvito,~S.; Hou,~S. Current-Induced Phonon Renormalization in Molecular
  Junctions. \emph{Phys. Rev. B} \textbf{2016}, \emph{94}, 035411\relax
\mciteBstWouldAddEndPuncttrue
\mciteSetBstMidEndSepPunct{\mcitedefaultmidpunct}
{\mcitedefaultendpunct}{\mcitedefaultseppunct}\relax
\EndOfBibitem
\bibitem[Frederiksen \latin{et~al.}(2007)Frederiksen, Paulsson, and
  Brandbyge]{frederiksenInelasticFingerprintsHydrogen2007}
Frederiksen,~T.; Paulsson,~M.; Brandbyge,~M. Inelastic Fingerprints of Hydrogen
  Contamination in Atomic Gold Wire Systems. \emph{J. Phys.: Conf. Ser.}
  \textbf{2007}, \emph{61}, 312--316\relax
\mciteBstWouldAddEndPuncttrue
\mciteSetBstMidEndSepPunct{\mcitedefaultmidpunct}
{\mcitedefaultendpunct}{\mcitedefaultseppunct}\relax
\EndOfBibitem
\bibitem[Yanson \latin{et~al.}(1998)Yanson, Bollinger, Van Den~Brom,
  Agra{\"i}t, and Van~Ruitenbeek]{yansonFormationManipulationMetallic1998}
Yanson,~A.~I.; Bollinger,~G.~R.; Van Den~Brom,~H.~E.; Agra{\"i}t,~N.;
  Van~Ruitenbeek,~J.~M. Formation and Manipulation of a Metallic Wire of Single
  Gold Atoms. \emph{Nature} \textbf{1998}, \emph{395}, 783--785\relax
\mciteBstWouldAddEndPuncttrue
\mciteSetBstMidEndSepPunct{\mcitedefaultmidpunct}
{\mcitedefaultendpunct}{\mcitedefaultseppunct}\relax
\EndOfBibitem
\bibitem[Untiedt \latin{et~al.}(2002)Untiedt, Yanson, Grande,
  {Rubio-Bollinger}, Agra{\"i}t, Vieira, and
  Van~Ruitenbeek]{untiedtCalibrationLengthChain2002}
Untiedt,~C.; Yanson,~A.~I.; Grande,~R.; {Rubio-Bollinger},~G.; Agra{\"i}t,~N.;
  Vieira,~S.; Van~Ruitenbeek,~J. Calibration of the Length of a Chain of Single
  Gold Atoms. \emph{Phys. Rev. B} \textbf{2002}, \emph{66}, 085418\relax
\mciteBstWouldAddEndPuncttrue
\mciteSetBstMidEndSepPunct{\mcitedefaultmidpunct}
{\mcitedefaultendpunct}{\mcitedefaultseppunct}\relax
\EndOfBibitem
\bibitem[{S{\'a}nchez-Portal} \latin{et~al.}(1999){S{\'a}nchez-Portal},
  Artacho, Junquera, Ordej{\'o}n, Garc{\'i}a, and
  Soler]{sanchez-portalStiffMonatomicGold1999}
{S{\'a}nchez-Portal},~D.; Artacho,~E.; Junquera,~J.; Ordej{\'o}n,~P.;
  Garc{\'i}a,~A.; Soler,~J.~M. Stiff {{Monatomic Gold Wires}} with a {{Spinning
  Zigzag Geometry}}. \emph{Phys. Rev. Lett.} \textbf{1999}, \emph{83},
  3884--3887\relax
\mciteBstWouldAddEndPuncttrue
\mciteSetBstMidEndSepPunct{\mcitedefaultmidpunct}
{\mcitedefaultendpunct}{\mcitedefaultseppunct}\relax
\EndOfBibitem
\bibitem[Torres \latin{et~al.}(1999)Torres, Tosatti, Dal~Corso, Ercolessi,
  Kohanoff, Di~Tolla, and Soler]{torresPuzzlingStabilityMonatomic1999}
Torres,~J.; Tosatti,~E.; Dal~Corso,~A.; Ercolessi,~F.; Kohanoff,~J.;
  Di~Tolla,~F.; Soler,~J. The Puzzling Stability of Monatomic Gold Wires.
  \emph{Surface Science} \textbf{1999}, \emph{426}, L441--L446\relax
\mciteBstWouldAddEndPuncttrue
\mciteSetBstMidEndSepPunct{\mcitedefaultmidpunct}
{\mcitedefaultendpunct}{\mcitedefaultseppunct}\relax
\EndOfBibitem
\bibitem[S{\o}rensen \latin{et~al.}(1998)S{\o}rensen, Brandbyge, and
  Jacobsen]{sorensenMechanicalDeformationAtomicscale1998}
S{\o}rensen,~M.~R.; Brandbyge,~M.; Jacobsen,~K.~W. Mechanical Deformation of
  Atomic-Scale Metallic Contacts: {{Structure}} and Mechanisms. \emph{Phys.
  Rev. B} \textbf{1998}, \emph{57}, 3283--3294\relax
\mciteBstWouldAddEndPuncttrue
\mciteSetBstMidEndSepPunct{\mcitedefaultmidpunct}
{\mcitedefaultendpunct}{\mcitedefaultseppunct}\relax
\EndOfBibitem
\bibitem[Menzel \latin{et~al.}(2015)Menzel, Kaupmann, and
  Waser]{menzelUnderstandingFilamentaryGrowth2015}
Menzel,~S.; Kaupmann,~P.; Waser,~R. Understanding Filamentary Growth in
  Electrochemical Metallization Memory Cells Using Kinetic {{Monte Carlo}}
  Simulations. \emph{Nanoscale} \textbf{2015}, \emph{7}, 12673--12681\relax
\mciteBstWouldAddEndPuncttrue
\mciteSetBstMidEndSepPunct{\mcitedefaultmidpunct}
{\mcitedefaultendpunct}{\mcitedefaultseppunct}\relax
\EndOfBibitem
\bibitem[Menzel(2014)]{menzelModelingSimulationElectrochemical2014}
Menzel,~S. Modeling and Simulation of Electrochemical Metallization Memory
  Cells. 2014 {{IEEE International Symposium}} on {{Circuits}} and {{Systems}}
  ({{ISCAS}}). Melbourne VIC, Australia, 2014; pp 2025--2028\relax
\mciteBstWouldAddEndPuncttrue
\mciteSetBstMidEndSepPunct{\mcitedefaultmidpunct}
{\mcitedefaultendpunct}{\mcitedefaultseppunct}\relax
\EndOfBibitem
\bibitem[Milano \latin{et~al.}(2019)Milano, Porro, Valov, and
  Ricciardi]{milanoRecentDevelopmentsPerspectives2019}
Milano,~G.; Porro,~S.; Valov,~I.; Ricciardi,~C. Recent {{Developments}} and
  {{Perspectives}} for {{Memristive Devices Based}} on {{Metal Oxide
  Nanowires}}. \emph{Adv Elect Materials} \textbf{2019}, \emph{5},
  1800909\relax
\mciteBstWouldAddEndPuncttrue
\mciteSetBstMidEndSepPunct{\mcitedefaultmidpunct}
{\mcitedefaultendpunct}{\mcitedefaultseppunct}\relax
\EndOfBibitem
\bibitem[Binnig \latin{et~al.}(1982)Binnig, Rohrer, Gerber, and
  Weibel]{binnigSurfaceStudiesScanning1982}
Binnig,~G.; Rohrer,~H.; Gerber,~{\relax Ch}.; Weibel,~E. Surface {{Studies}} by
  {{Scanning Tunneling Microscopy}}. \emph{Phys. Rev. Lett.} \textbf{1982},
  \emph{49}, 57--61\relax
\mciteBstWouldAddEndPuncttrue
\mciteSetBstMidEndSepPunct{\mcitedefaultmidpunct}
{\mcitedefaultendpunct}{\mcitedefaultseppunct}\relax
\EndOfBibitem
\bibitem[Muller \latin{et~al.}(1992)Muller, Van~Ruitenbeek, and
  De~Jongh]{mullerConductanceSupercurrentDiscontinuities1992}
Muller,~C.~J.; Van~Ruitenbeek,~J.~M.; De~Jongh,~L.~J. Conductance and
  Supercurrent Discontinuities in Atomic-Scale Metallic Constrictions of
  Variable Width. \emph{Phys. Rev. Lett.} \textbf{1992}, \emph{69},
  140--143\relax
\mciteBstWouldAddEndPuncttrue
\mciteSetBstMidEndSepPunct{\mcitedefaultmidpunct}
{\mcitedefaultendpunct}{\mcitedefaultseppunct}\relax
\EndOfBibitem
\bibitem[Wang \latin{et~al.}(2019)Wang, Wang, Zhang, and
  Xiang]{wangAdvanceMechanicallyControllable2019}
Wang,~L.; Wang,~L.; Zhang,~L.; Xiang,~D. In \emph{Molecular-{{Scale
  Electronics}}}; Guo,~X., Ed.; Springer International Publishing: Cham, 2019;
  pp 45--86\relax
\mciteBstWouldAddEndPuncttrue
\mciteSetBstMidEndSepPunct{\mcitedefaultmidpunct}
{\mcitedefaultendpunct}{\mcitedefaultseppunct}\relax
\EndOfBibitem
\bibitem[Jorgensen \latin{et~al.}(1983)Jorgensen, Chandrasekhar, Madura, Impey,
  and Klein]{jorgensenComparisonSimplePotential1983a}
Jorgensen,~W.~L.; Chandrasekhar,~J.; Madura,~J.~D.; Impey,~R.~W.; Klein,~M.~L.
  Comparison of Simple Potential Functions for Simulating Liquid Water.
  \emph{The Journal of Chemical Physics} \textbf{1983}, \emph{79},
  926--935\relax
\mciteBstWouldAddEndPuncttrue
\mciteSetBstMidEndSepPunct{\mcitedefaultmidpunct}
{\mcitedefaultendpunct}{\mcitedefaultseppunct}\relax
\EndOfBibitem
\bibitem[Berg \latin{et~al.}(2017)Berg, Peter, and
  Johnston]{bergEvaluationOptimizationInterface2017}
Berg,~A.; Peter,~C.; Johnston,~K. Evaluation and {{Optimization}} of
  {{Interface Force Fields}} for {{Water}} on {{Gold Surfaces}}. \emph{J. Chem.
  Theory Comput.} \textbf{2017}, \emph{13}, 5610--5623\relax
\mciteBstWouldAddEndPuncttrue
\mciteSetBstMidEndSepPunct{\mcitedefaultmidpunct}
{\mcitedefaultendpunct}{\mcitedefaultseppunct}\relax
\EndOfBibitem
\bibitem[Odell \latin{et~al.}(2010)Odell, Delin, Johansson, Rungger, and
  Sanvito]{odellInvestigationConductingProperties2010}
Odell,~A.; Delin,~A.; Johansson,~B.; Rungger,~I.; Sanvito,~S. Investigation of
  the {{Conducting Properties}} of a {{Photoswitching Dithienylethene
  Molecule}}. \emph{ACS Nano} \textbf{2010}, \emph{4}, 2635--2642\relax
\mciteBstWouldAddEndPuncttrue
\mciteSetBstMidEndSepPunct{\mcitedefaultmidpunct}
{\mcitedefaultendpunct}{\mcitedefaultseppunct}\relax
\EndOfBibitem
\bibitem[Toher and Sanvito(2008)Toher, and
  Sanvito]{toherEffectsSelfinteractionCorrections2008}
Toher,~C.; Sanvito,~S. Effects of Self-Interaction Corrections on the Transport
  Properties of Phenyl-Based Molecular Junctions. \emph{Phys. Rev. B}
  \textbf{2008}, \emph{77}, 155402\relax
\mciteBstWouldAddEndPuncttrue
\mciteSetBstMidEndSepPunct{\mcitedefaultmidpunct}
{\mcitedefaultendpunct}{\mcitedefaultseppunct}\relax
\EndOfBibitem
\bibitem[Henkelman \latin{et~al.}(2006)Henkelman, Arnaldsson, and
  J{\'o}nsson]{Henkelman2006}
Henkelman,~G.; Arnaldsson,~A.; J{\'o}nsson,~H. A Fast and Robust Algorithm for
  {{Bader}} Decomposition of Charge Density. \emph{Computational Materials
  Science} \textbf{2006}, \emph{36}, 354--360\relax
\mciteBstWouldAddEndPuncttrue
\mciteSetBstMidEndSepPunct{\mcitedefaultmidpunct}
{\mcitedefaultendpunct}{\mcitedefaultseppunct}\relax
\EndOfBibitem
\bibitem[Spellacy \latin{et~al.}(2019)Spellacy, Golden, and
  Rungger]{spellacyPerformanceAnalysisPairwise2019}
Spellacy,~L.; Golden,~D.; Rungger,~I. Performance Analysis of a Pairwise Method
  for Partial Inversion of Complex Block Tridiagonal Matrices.
  \emph{Concurrency Computat Pract Exper} \textbf{2019}, \emph{31}\relax
\mciteBstWouldAddEndPuncttrue
\mciteSetBstMidEndSepPunct{\mcitedefaultmidpunct}
{\mcitedefaultendpunct}{\mcitedefaultseppunct}\relax
\EndOfBibitem
\bibitem[Behler(2021)]{Behler2021}
Behler,~J. Four {{Generations}} of {{High-Dimensional Neural Network
  Potentials}}. \emph{Chemical Reviews} \textbf{2021}, \emph{121},
  10037--10072\relax
\mciteBstWouldAddEndPuncttrue
\mciteSetBstMidEndSepPunct{\mcitedefaultmidpunct}
{\mcitedefaultendpunct}{\mcitedefaultseppunct}\relax
\EndOfBibitem
\bibitem[Zhou \latin{et~al.}(2023)Zhou, Ouyang, Zhang, Li, and
  Wang]{zhouMachineLearningAssisted2023}
Zhou,~Y.; Ouyang,~Y.; Zhang,~Y.; Li,~Q.; Wang,~J. Machine {{Learning Assisted
  Simulations}} of {{Electrochemical Interfaces}}: {{Recent Progress}} and
  {{Challenges}}. \emph{J. Phys. Chem. Lett.} \textbf{2023}, \emph{14},
  2308--2316\relax
\mciteBstWouldAddEndPuncttrue
\mciteSetBstMidEndSepPunct{\mcitedefaultmidpunct}
{\mcitedefaultendpunct}{\mcitedefaultseppunct}\relax
\EndOfBibitem
\end{mcitethebibliography}
\providecommand{\latin}[1]{#1}
\makeatletter
\providecommand{\doi}
  {\begingroup\let\do\@makeother\dospecials
  \catcode`\{=1 \catcode`\}=2 \doi@aux}
\providecommand{\doi@aux}[1]{\endgroup\texttt{#1}}
\makeatother
\providecommand*\mcitethebibliography{\thebibliography}
\csname @ifundefined\endcsname{endmcitethebibliography}
  {\let\endmcitethebibliography\endthebibliography}{}

\end{document}